\documentclass[aos]{imsart}

\RequirePackage{amsthm,amsmath,amsfonts,amssymb}
\RequirePackage[authoryear]{natbib}
\RequirePackage[colorlinks,citecolor=blue,urlcolor=blue]{hyperref}
\RequirePackage{graphicx}
\usepackage{algorithm}
\usepackage{algpseudocode}
\usepackage{subcaption}
\usepackage{dsfont}
\usepackage{booktabs}
\usepackage{pifont}

\usepackage{longtable}
\usepackage{xcolor}
\startlocaldefs
\numberwithin{equation}{section}
\theoremstyle{plain}

\newtheorem{theorem}{Theorem}[section]

\theoremstyle{definition}
\newtheorem{remark}{Remark}[section]

\newtheorem{example}{Example}[section]

\endlocaldefs

\begin{document}

\begin{frontmatter}
\title{Minimax and Adaptive Transfer Learning for Sparse Canonical Correlation Analysis}
\runtitle{Trans sCCA}

\begin{aug}
\author[A]{\fnms{Bo}~\snm{Fu}\ead[label=e1]{311342@stu.xjtu.edu.cn}}
\and
\author[A]{\fnms{Dandan}~\snm{Jiang}\thanks{[\textbf{Corresponding author.}]}\ead[label=e2]{jiangdd@xjtu.edu.cn}}
\address[A]{School of Mathematics and Statistics, Xi'an Jiaotong University\printead[presep={ ,\ }]{e1,e2}}
\end{aug}

\begin{abstract}
We develop a transfer learning framework for high-dimensional sparse canonical correlation analysis (CCA) with multiple heterogeneous source datasets. Our goal is to improve estimation of the target canonical coefficient matrices by borrowing information from related sources. We characterize source-target similarity through discrepancies in the joint covariance structure and establish minimax optimal rates under both joint and separate prediction losses, explicitly quantifying the gain from enlarged effective sample size and the cost of heterogeneity. To attain these rates, we first construct an oracle transfer estimator based on debiasing and truncated normalization. We then develop a computationally efficient procedure that adaptively achieves the minimax rate without prior knowledge of the sparsity levels, using aggregation to control transfer bias and a truncated normalization scheme to preserve transfer gains under the scale-sensitive prediction loss. A data-driven source detection procedure is further proposed to identify potentially informative auxiliary datasets. Our analysis also yields a generalized sin-theta perturbation theorem that removes a commonly imposed comparability condition on leading canonical correlations in the sparse CCA problem. 
\end{abstract}

\begin{keyword}[class=MSC]
\kwd[Primary ]{62H12}
\kwd[; secondary ]{62C20}
\end{keyword}

\begin{keyword}
\kwd{Canonical Correlation Analysis}
\kwd{Transfer Learning}
\kwd{Minimax Rates}
\kwd{Covariance Matrix}
\kwd{Optimality}
\end{keyword}

\end{frontmatter}

\section{Introduction}\label{sec1}
Modern scientific studies increasingly collect data from multiple related populations, cohorts, or sites. In many such applications, the target population of primary scientific interest may have only a limited sample size, while auxiliary datasets from related populations are available and may contain useful information about the target population. 

Classical Canonical Correlation Analysis (CCA; \citealp{hotelling1936relations}) is a fundamental tool in multivariate analysis for characterizing the association between two sets of variables through maximally correlated linear combinations. It has found important applications in neuroimaging, imaging genetics, and multi-omics studies \citep{butler2018integrating,jiang2023canonical}. However, it is commonly encountered in practice that the data dimensions can be larger than the sample size \citep{bai2022limiting,bao2019canonical,gao2015minimax,gao2017sparse}, in which classical CCA breaks down, and the leading canonical coefficient matrices are commonly assumed to be row sparse \citep{waaijenborg2009sparse,hardoon2011sparse,le2009sparse,parkhomenko2009sparse,gao2015minimax,gao2017sparse,witten2009penalized} to improve estimation and interpretability, which is called sparse CCA.

Most existing theory and methodology for classical CCA and sparse CCA, nevertheless, have been developed for a single dataset, which can be restrictive in modern applications where the target sample size is limited but related auxiliary datasets are available. In practice, classical CCA and sparse CCA can exhibit substantial instability in small samples \citep{helmer2024stability}, while modern genomic, neuroimaging, and multimodal studies are often conducted across related cohorts, hospitals, or sites that may share common cross-view association structures. For example, \cite{jiang2023canonical} applied sparse CCA to proteomic and methylomic data from two independent cohorts and found that canonical variables learned from one cohort retained explanatory power in another, providing empirical evidence for the transferability of canonical structures. 

Despite this potential transferability, directly pooling related datasets may lead to biased estimation when their underlying distributions are heterogeneous. Transfer learning \citep{pan2009survey} has become a powerful tool for leveraging auxiliary datasets to improve performance while accounting for such heterogeneity. A common message from this literature is that auxiliary samples can substantially improve estimation when source and target populations are sufficiently similar. It is therefore natural to ask whether related source datasets can be used to improve sparse CCA estimation for a target population.

\subsection{Related work}
In the conventional single-dataset setting, sparse CCA has been extensively studied from both methodological and theoretical perspectives \citep{witten2009penalized,gao2015minimax,gao2017sparse,mai2019iterative}. More recent work has extended sparse CCA to broader distributional settings, including mixed data through latent Gaussian copula models \citep{yoon2020sparse} and heavy-tailed data through sparse Gaussianized CCA \citep{di2026sparse}. These methods, however, primarily focus on estimating canonical structures from a single population.

Several extensions of CCA have considered distributed learning or federated learning under differential privacy \citep{7120160,chen2024distributed,luo2026fedcca}. These methods primarily aim to estimate a common canonical structure from decentralized data, rather than selectively exploiting heterogeneous auxiliary datasets to improve estimation for a prespecified target population.

In parallel, transfer learning has been extensively studied in a variety of statistical problems, including nonparametric classification \citep{cai2021transfer}, transfer reinforcement learning \citep{chai2025deep}, principal component analysis (PCA; \citealp{li2024knowledge}), covariance matrix estimation \citep{cai2026transfer}, precision matrix estimation \citep{li2022transfer,zhao2026trans}, and parametric regression models \citep{li2022transfer,bastani2021predicting,tian2023transfer,jin2024transfer,fu2025transfer,li2024estimation}. Despite these advances, to the best of our knowledge, the statistical problem of target-specific sparse CCA estimation with heterogeneous auxiliary populations and its fundamental minimax limits, has not been systematically studied.

\subsection{The Problem under Study and Challenges in Transfer learning for Sparse CCA}\label{sec1.1}
In this work, we are interested in enhancing the estimation of sparse CCA from available auxiliary datasets. Roughly speaking, suppose that we have a target dataset $\{\boldsymbol{x}_i^{(0)},\boldsymbol{y}_i^{(0)}\}_{i=1}^{n^{(0)}}$ and $K$ independent source datasets $\left\{\{\boldsymbol{x}_i^{(k)},\boldsymbol{y}_i^{(k)}\}_{i=1}^{n^{(k)}}\right\}_{k=1}^{K}$, where $\boldsymbol{x}_i^{(k)}\in\mathbb{R}^{p}$ and $\boldsymbol{y}_i^{(k)}\in\mathbb{R}^{m}$. Without loss of generality, we assume that $\boldsymbol{x}_i^{(k)}$ and $\boldsymbol{y}_i^{(k)}$ are centered, i.e., with zero-mean. Let $\boldsymbol{\Sigma}^{(k)}$ be the covariance matrix of $(\boldsymbol{x}_i^{(k)},\boldsymbol{y}_i^{(k)})$, where
\begin{align}\label{eq1.1}
    \boldsymbol{\Sigma}^{(k)}=\begin{bmatrix}
    \boldsymbol{\Sigma}_x^{(k)} & \boldsymbol{\Sigma}_{xy}^{(k)} \\
    \boldsymbol{\Sigma}_{yx}^{(k)} & \boldsymbol{\Sigma}_y^{(k)} 
\end{bmatrix}.
\end{align}
At the population level, CCA for the target dataset searches successive vectors $\boldsymbol{u}_j^{(0)}\in\mathbb{R}^{p}$ and $\boldsymbol{v}_j^{(0)}\in\mathbb{R}^{m}$ such that
\begin{equation}\label{eq1.2}
    \begin{aligned}
        &\boldsymbol{u}_j^{(0)},\boldsymbol{v}_j^{(0)}=\mathop{\arg\max}_{\boldsymbol{u},\boldsymbol{v}}\boldsymbol{u}^\top \boldsymbol{\Sigma}_{xy}^{(0)} \boldsymbol{v},\\
        &\text{s.t.}~\boldsymbol{u}^\top \boldsymbol{\Sigma}_{x}^{(0)} \boldsymbol{u}=\boldsymbol{v}^\top \boldsymbol{\Sigma}_{y}^{(0)} \boldsymbol{v}=1,\,\boldsymbol{u}^\top \boldsymbol{\Sigma}_{x}^{(0)} \boldsymbol{u}_{l}^{(0)}=\boldsymbol{v}^\top \boldsymbol{\Sigma}_{y}^{(0)} \boldsymbol{v}_{l}^{(0)}=0,\forall 0\leq l\leq j-1
    \end{aligned}
\end{equation}
where we write $\boldsymbol{u}_{0}^{(0)}=\boldsymbol{0}_{p}$ and $\boldsymbol{v}_{0}^{(0)}=\boldsymbol{0}_{m}$. The $j$-th pair canonical variates is then the linear combinations $(\boldsymbol{u}_j^{(0)\top}\boldsymbol{x}_i^{(0)},\boldsymbol{v}_j^{(0)\top}\boldsymbol{y}_i^{(0)})$. Following \cite{chen2013sparse,gao2017sparse}, we assume that the target cross-covariance matrix has $r$ nonzero canonical correlations and hence admits $r$ canonical pairs. Then, $\boldsymbol{\Sigma}_{xy}^{(0)}$ in the form of \eqref{eq1.2} can be reparameterized as 
\begin{align}\label{eq1.3}
    \boldsymbol{\Sigma}_{xy}^{(0)}=\boldsymbol{\Sigma}_{x}^{(0)}\boldsymbol{U}^{(0)}\boldsymbol{\Lambda}^{(0)}\boldsymbol{V}^{(0)\top}\boldsymbol{\Sigma}_{y}^{(0)},\quad\text{where}\quad \boldsymbol{U}^{(0)\top}\boldsymbol{\Sigma}_{x}^{(0)}\boldsymbol{U}^{(0)}=\boldsymbol{V}^{(0)\top}\boldsymbol{\Sigma}_{y}^{(0)}\boldsymbol{V}^{(0)}=\boldsymbol{I}_{r}.
\end{align}
Here, $\boldsymbol{U}^{(0)}=[\boldsymbol{u}_1^{(0)},\ldots,\boldsymbol{u}_r^{(0)}]$ and $\boldsymbol{V}^{(0)}=[\boldsymbol{v}_1^{(0)},\ldots,\boldsymbol{v}_r^{(0)}]$ are the canonical coefficient vectors, and $\boldsymbol{\Lambda}^{(0)}=\mbox{diag}(\lambda_1^{(0)},\ldots,\lambda_r^{(0)})$ is a diagonal matrix with $1>\lambda_1^{(0)}\geq\ldots\geq\lambda_r^{(0)}\geq\lambda>0$ being the descending canonical correlations. When only the target dataset is available, under a Gaussian canonical pair model, \cite{gao2015minimax} established the minimax rate for jointly estimating $\boldsymbol{U}^{(0)}$ and $\boldsymbol{V}^{(0)}$ under a specified loss function $\|\hat{\boldsymbol{U}}\hat{\boldsymbol{V}}^\top-\boldsymbol{U}^{(0)}\boldsymbol{V}^{(0)\top}\|_{\mathrm{F}}^2$ is $$\frac{1}{n^{(0)}\lambda^2}\left(r(s_u+s_v)+s_u\log\left(\frac{ep}{s_u}\right)+s_v\log\left(\frac{em}{s_v}\right)\right)$$
by assuming that $|\mathrm{supp}(\boldsymbol{U}^{(0)})|\leq s_u$ and $|\mathrm{supp}(\boldsymbol{V}^{(0)})|\leq s_v$, where $\mathrm{supp}(\boldsymbol{A})$ denotes the nonzero rows for a specified matrix $\boldsymbol{A}$. \cite{gao2017sparse} established separate minimax rates for $\boldsymbol{U}^{(0)}$ under the loss function (see also in Equation \eqref{eq2.5})
\begin{align}\label{eq1.4}
    L(\hat{\boldsymbol{U}},\boldsymbol{U}^{(0)})=\inf_{\boldsymbol{W}\in\mathcal{O}(r,r)}\mathbb{E}_{\boldsymbol{x}'\mathop{=}\limits^{d}\boldsymbol{x}^{(0)}\sim\mathbb{N}(\boldsymbol{0}_p,\boldsymbol{\Sigma}_{x}^{(0)})}\|\boldsymbol{W}^\top\hat{\boldsymbol{U}}^{\top}\boldsymbol{x}'-\boldsymbol{U}^{(0)\top}\boldsymbol{x}'\|_2^2
\end{align}
is 
$$\frac{1}{n^{(0)}\lambda^2}\left(s_ur+s_u\log\left(\frac{ep}{s_u}\right)\right),$$
where $\mathcal{O}(r,r)$ denotes the collection of all $r\times r$ orthogonal matrices. Following \cite{gao2017sparse}, we use this prediction loss to measure how accurately the estimated canonical variates recover their population counterparts for a new target observation, up to an orthogonal transformation. It captures errors in both the estimated canonical directions and their normalization, hence containing more information than the projection loss.


When auxiliary source datasets are available, it is expected that leveraging these datasets will enhance performance compared to employing the target dataset only. Our goal is to construct a computationally efficient and minimax optimal estimator for $\boldsymbol{U}^{(0)}$ under the loss function \eqref{eq1.4} (and symmetrically, for $\boldsymbol{V}^{(0)}$). However, transfer learning for CCA is intrinsically more delicate than for problems involving a single covariance matrix. Unlike PCA, where the principal subspace is determined by a single covariance matrix and is orthogonal under Euclidean geometry, the canonical subspaces in CCA depend on both marginal and cross-covariance matrices and are normalized for unknown marginal covariance structures. This additional dependency makes CCA more challenging; see \cite{gao2015minimax} for a detailed comparison of sparse CCA and sparse PCA. Consequently, heterogeneity across datasets can affect the target canonical directions through the two marginal covariance matrices and the cross-covariance matrix. Naively pooling source and target data can therefore introduce substantial transfer bias. 

The goal of the present work is to address the following questions: (1) How should similarity across datasets be quantified, and what are the corresponding minimax rates for estimating $\boldsymbol{U}^{(0)}$ and $\boldsymbol{V}^{(0)}$ under the joint and the separate losses? (2) Is there any computationally efficient algorithm that accommodates source-target heterogeneity and adaptively achieves the minimax rate without prior knowledge of $s_u$ and $s_v$? If so, what is the price of such adaptivity? (3) How can the canonical coefficient vectors be normalized with respect to the unknown target population covariance matrices while preserving the gains from transfer under the prediction loss? 

To illustrate the difficulty underlying the last question, consider the following toy example.
\begin{example}\label{example 1}
    Suppose that $p=m=2$ and $r=1$. We consider the joint Gaussian model $(\boldsymbol{x}_i^{(k)},\boldsymbol{y}_i^{(k)})\sim\mathbb{N}(\boldsymbol{0}_{4},\boldsymbol{\Sigma}^{(k)})$ with $$\boldsymbol{\Sigma}^{(k)}=\boldsymbol{\Sigma}=\begin{bmatrix}
    \boldsymbol{I}_2 & \lambda\boldsymbol{e}_1\boldsymbol{e}_1^\top \\
    \lambda\boldsymbol{e}_1\boldsymbol{e}_1^\top & \boldsymbol{I}_2 
    \end{bmatrix},\quad\lambda\in(0,1),\quad\forall k\in\{0,1,\ldots,K\}$$
    where $\boldsymbol{e}_1=(1,0)^\top$. In this case, $\boldsymbol{U}^{(0)}=\boldsymbol{V}^{(0)}=\boldsymbol{e}_1$, and no heterogeneity exists across datasets. To isolate the effect of normalization, suppose, in the most favorable case, that the estimator exactly recovers the true canonical coefficient vector, i.e., $\hat{\boldsymbol{U}}=\boldsymbol{U}^{(0)}=\boldsymbol{e}_1$. Denote $\hat{\boldsymbol{\Sigma}}_x^{(0)}=\sum_{i=1}^{n^{(0)}}\boldsymbol{x}_i^{(0)}\boldsymbol{x}_i^{(0)\top}/n^{(0)}$ as the sample covariance matrix of $\boldsymbol{x}_i^{(0)}$ with sample size $n^{(0)}$. To normalize $\hat{\boldsymbol{U}}$ with respect to $\hat{\boldsymbol{\Sigma}}_x^{(0)}$, define $\tilde{\boldsymbol{U}}=\hat{\boldsymbol{U}}\left(\hat{\boldsymbol{U}}^\top\hat{\boldsymbol{\Sigma}}_x^{(0)}\hat{\boldsymbol{U}}\right)^{-1/2}$ (if invertible) such that $\tilde{\boldsymbol{U}}^\top\hat{\boldsymbol{\Sigma}}_x^{(0)}\tilde{\boldsymbol{U}}=\boldsymbol{I}_r=1$. Then the loss in \eqref{eq1.4} can be calculated as 
    \begin{align*}
        L(\tilde{\boldsymbol{U}},\boldsymbol{U}^{(0)})=&\inf_{w\in\{-1,1\}}\mathbb{E}_{\boldsymbol{x}'\sim\mathbb{N}(\boldsymbol{0}_2,\boldsymbol{I}_2)}\|w\tilde{\boldsymbol{U}}^{\top}\boldsymbol{x}'-\boldsymbol{U}^{(0)\top}\boldsymbol{x}'\|_2^2\\
        =&\inf_{w\in\{-1,1\}}\|w\tilde{\boldsymbol{U}}^{\top}-\boldsymbol{U}^{(0)\top}\|_2^2=\inf_{w\in\{-1,1\}}\big\|w\hat{\boldsymbol{U}}^\top\big(\hat{\boldsymbol{U}}^\top\hat{\boldsymbol{\Sigma}}_x^{(0)}\hat{\boldsymbol{U}}\big)^{-1/2}-\boldsymbol{U}^{(0)\top}\big\|_2^2\\
        =&\left(\big(\hat{\boldsymbol{U}}^\top\hat{\boldsymbol{\Sigma}}_x^{(0)}\hat{\boldsymbol{U}}\big)^{-1/2}-1\right)^2.
    \end{align*}
    Since $n^{(0)}\hat{\boldsymbol{U}}^\top\hat{\boldsymbol{\Sigma}}_x^{(0)}\hat{\boldsymbol{U}}\sim\chi^2_{n^{(0)}}$, the central limit theorem and the delta method give that $(n^{(0)})^{1/2}\big[(\hat{\boldsymbol{U}}^{\top}\hat{\boldsymbol{\Sigma}}_{x}^{(0)}\hat{\boldsymbol{U}})^{-1/2}-1\big]\xrightarrow{d}N(0,1/2)$. Thus, even when the population coefficient vector is recovered exactly, enforcing exact normalization with respect to the target sample covariance matrix introduces an additional error of order $1/n^{(0)}$ under the prediction loss \eqref{eq1.4}. This error can dominate the pooled estimation error in favorable transfer regimes. The same phenomenon can be expressed directly in terms of population normalization. Since $\boldsymbol{\Sigma}_x^{(0)}=\boldsymbol{I}_2$, we have $|\tilde{\boldsymbol{U}}^\top\boldsymbol{\Sigma}_x^{(0)}\tilde{\boldsymbol{U}}-1|=|(\hat{\boldsymbol{U}}^\top\hat{\boldsymbol{\Sigma}}_x^{(0)}\hat{\boldsymbol{U}})^{-1}-1|$, which is of order $(n^{(0)})^{-1/2}$. Therefore, the example isolates a discrepancy between exact sample normalization and accurate estimation of the population-normalized canonical coefficient vector: the former is natural for computing canonical correlations, but it need not be optimal for coefficient estimation under the scale-sensitive prediction loss \eqref{eq1.4}. Note that $|\tilde{\boldsymbol{U}}^\top\boldsymbol{\Sigma}_x^{(0)}\tilde{\boldsymbol{U}}-1|$ here quantifies the population-level normalization error. This motivates the population-level normalization error introduced in \eqref{eq2.6}. The estimator of \cite{gao2017sparse} likewise imposes exact normalization with respect to a target sample covariance matrix and is therefore subject to a population normalization bound limited by target sample size.
\end{example}

\subsection{Main Results and Our Contributions}\label{sec1.3}
In this paper, we propose a \textbf{Trans}fer learning procedure for \textbf{s}parse \textbf{CCA} (Trans-sCCA). We characterize source-target similarity through the joint covariance matrix. Specifically, we regard the $k$-th source dataset as informative if $\|\boldsymbol{\Sigma}^{(k)}-\boldsymbol{\Sigma}^{(0)}\|_2\leq h$, where $\boldsymbol{\Sigma}^{(0)}$ and $\boldsymbol{\Sigma}^{(k)}$ denote the joint population covariance matrices of target and the $k$-th source datasets, respectively, and $\|\cdot\|_2$ represents the spectral norm. Similar conditions are required by \cite{cai2026transfer} for covariance matrix estimation. Moreover, this condition can be weakened to some extent (see Remark \ref{remark 1}) and does not affect the results in this paper. This condition controls for the differences between the two marginal and cross covariance matrices. Let $\mathcal{A}_h=\big\{k\in[K]: \|\boldsymbol{\Sigma}^{(k)}-\boldsymbol{\Sigma}^{(0)}\|_2\leq h\big\}$. Under this similarity measure, we show that the separate minimax rate for $\boldsymbol{U}^{(0)}$ under the loss function \eqref{eq1.4} is 
$$\frac{s_ur+s_u\log(ep/s_u)}{(n^{(0)}+n^{\mathcal{A}_h})\lambda^2}+\frac{rh^2}{\lambda^2}\wedge\frac{s_ur+s_u\log(ep/s_u)}{n^{(0)}\lambda^2},$$
by assuming that $|\mathrm{supp}(\boldsymbol{U}^{(0)})|\leq s_u,~|\mathrm{supp}(\boldsymbol{V}^{(0)})|\leq s_v$, where $n^{\mathcal{A}_h}=\sum_{k\in\mathcal{A}_h}n^{(k)}$ is the sample size of the informative dataset. As a by-product, we also show that the minimax rate for jointly estimating $\boldsymbol{U}^{(0)}$ and $\boldsymbol{V}^{(0)}$ under the loss function $\|\hat{\boldsymbol{U}}\hat{\boldsymbol{V}}^\top-\boldsymbol{U}^{(0)}\boldsymbol{V}^{(0)\top}\|_{\mathrm{F}}^2$ is $$\frac{\mathcal{C}_{u,v}}{(n^{(0)}+n^{\mathcal{A}_h})\lambda^2}+\frac{rh^2}{\lambda^2}\wedge\frac{\mathcal{C}_{u,v}}{n^{(0)}\lambda^2},$$
where $\mathcal{C}_{u,v}=r(s_u+s_v)+s_u\log(ep/s_u)+s_v\log(em/s_v)$. In these two rates, the first terms capture the gain from the enlarged effective sample size, whereas the second terms quantify the cost of source-target heterogeneity.

To attain the separate minimax rate for $\boldsymbol{U}^{(0)}$, two key techniques are required. First, the transfer bias induced by pooling must be corrected without losing the improvement provided by the source datasets. We propose debiasing and aggregation approaches for this purpose. Second, the resulting estimator must be normalized without reintroducing an error determined solely by the target sample size. As illustrated in Example \ref{example 1}, exact normalization with respect to the target sample covariance matrix can reintroduce additional stochastic error with a rate limited by target sample size, while normalization with respect to the pooled covariance matrix using all data may retain bias caused by source-target heterogeneity. We therefore develop a truncated normalization scheme that balances these two effects. Our truncated normalization preserves the gains from transfer under the scale-sensitive prediction loss while also achieving smaller population-level normalization loss \eqref{eq2.6}. This truncated normalization also improves numerical stability by avoiding a potentially unstable matrix inversion when the target sample size is small relative to $r$. In conclusion, our contributions are summarized as follows:
\begin{itemize}
    \item We develop a transfer learning framework for sparse CCA that leverages auxiliary source datasets to enhance the estimation accuracy and establish the minimax rates under both joint and separate loss functions.
    \item We propose an oracle transfer sparse CCA estimator (Algorithm \ref{algo1}) that employs transfer, a debiasing technique under $\ell_0$-constraint, and a truncated normalization scheme and achieves the minimax-optimal rate under the separate loss function \eqref{eq1.4} when the set of informative source datasets and the sparsity levels are known. We further propose a computationally efficient estimator (Algorithm \ref{algo2}) when the informative source set is known, which, under additional sample size conditions, adaptively matches the minimax-optimal rate without knowledge of the sparsity levels, through an aggregation step and the same normalization procedure. Both estimators also yield smaller population-level normalization error. To prevent negative transfer, we apply a source detection algorithm. 
    \item We provide a more general version of the generalized sin-theta theorem that does not require the leading singular values to be comparable to the smallest signal singular value imposed in Theorem 5 of \cite{gao2015minimax}; see Theorem \ref{TheoremB: Generalized sin-theta theorem without condition of same order singular values}. In the target-only CCA setting, this in particular removes the requirement $\lambda_1^{(0)}\leq C\lambda_r^{(0)}$ imposed in \cite{gao2015minimax}. Beyond the present application, this result provides a useful perturbation bound for problems in which the leading singular values need not be comparable to the smallest signal singular value and may be of interest in its own right.
\end{itemize}

\subsection{Roadmap and Notation}
The remainder of this paper is structured as follows. Section \ref{sec2} formulates the statistical problem including the parameter space, the similarity measure between source and target domains, and the loss function. Section \ref{sec3} investigates the fundamental statistical limits of transfer sparse CCA and constructs an oracle estimator that achieves the minimax optimal convergence rate. Section \ref{sec4} develops computationally feasible and adaptive methods and proposes a data-driven approach to identify potentially useful source datasets. Section \ref{sec5} presents numerical results and a real data analysis. Section \ref{sec6} concludes the paper. All proofs and additional simulation results for data-driven source detection are provided in the Supplementary Material.

\noindent
\textbf{Notation:} For a vector $\boldsymbol{v}=(v_1,\ldots,v_p)^\top\in\mathbb{R}^{p}$, denote $\|\boldsymbol{v}\|_2=(\sum_{i=1}^{p}|v_i|^2)^{1/2}$, $\|\boldsymbol{v}\|_1=\sum_{i=1}^{p}|v_i|$, $\|\boldsymbol{v}\|_\infty=\max_{1\leq i\leq p}|v_i|$. Denote $[p]=\{1,\ldots,p\}$. We use $\mathrm{supp}(\boldsymbol{v})=\{j\in[p]:v_j\neq 0\}$ to indicate the support set of $\boldsymbol{v}$. We write $[\boldsymbol{v}]_{S}$ as the sub-vector of $\boldsymbol{v}$ for any $S\subseteq[p]$ and abbreviate it as $\boldsymbol{v}_{S}$ when there is no ambiguity. Let $\nabla_{S}f(\boldsymbol{v})$ be the gradient of $f(\boldsymbol{v})$ with respect to $\boldsymbol{v}_{S}$ for a given function $f$. Denote $\mathbb{S}^{l-1}$ as the $(l-1)$-dimensional unit sphere in $\mathbb{R}^{l}$. For a matrix $\boldsymbol{A}=(A_{i,j})_{1\leq i\leq k,1\leq j\leq l}\in \mathbb{R}^{k\times l}$, denote $\|\boldsymbol{A}\|_2=\max_{\boldsymbol{v}\in\mathbb{S}^{l-1}}\|\boldsymbol{A}\boldsymbol{v}\|_2$. Write $\|\boldsymbol{v}\|_{\boldsymbol{A}}=\|\boldsymbol{A}^{1/2}\boldsymbol{v}\|_2$ to denote the vector norm induced by $\boldsymbol{A}$. Denote $a\vee b=\max\{a,b\}$ and $a\wedge b=\min\{a,b\}$. Let $\sigma_i(\boldsymbol{A})$ be the $i$-th singular value of $\boldsymbol{A}$ and $\lambda_i(\boldsymbol{A})$ for a square matrix $\boldsymbol{A}$ be the $i$-th eigenvalue of $\boldsymbol{A}$, and we use $\sigma_{\min}(\boldsymbol{A})=\sigma_{k\wedge l}(\boldsymbol{A})$ and $\lambda_{\min}(\boldsymbol{A})=\lambda_{k}(\boldsymbol{A})$. We use $\boldsymbol{A}_{S,T}=[A_{i,j}]_{i\in S,j\in T}$ to represent the sub-matrix of $\boldsymbol{A}$ and abbreviate it as $\boldsymbol{A}_{S}$ if $S=T$. We further let $\|\boldsymbol{A}\|_{\mathrm{F}}=\sqrt{\sum_{i\in[k],j\in[l]}A_{i,j}^2},~\|\boldsymbol{A}\|_*=\sum_{i=1}^{k\wedge l}\sigma_i(\boldsymbol{A}),~\|\boldsymbol{A}\|_1=\sum_{i\in[k],j\in[l]}|A_{i,j}|,~\|\boldsymbol{A}\|_{2,1}=\sum_{i\in[k]}\|A_{i,[l]}\|_2,~\|\boldsymbol{A}\|_{2,\infty}=\max_{i\in[k]}\|A_{i,[l]}\|_2,~\|\boldsymbol{A}\|_{\max}=\max_{i\in[k],j\in[l]}|A_{i,j}|$ be the Frobenius norm, the nuclear norm, the $\ell_1$ norm, the $\ell_{2,1}$ norm, the maximum row $\ell_2$ norm, and the element-wise max-norm, respectively. For two sequences of non-negative numbers $\{a_n\}$ and $\{b_n\}$, we write $a_n\lesssim b_n$ if there exists a constant $C>0$ independent of $n$ such that $a_n\leq C b_n$. Similarly, $a_n\gtrsim b_n$ means that $b_n\lesssim a_n$. We write 
$\mathcal{O}(p,r)=\{\boldsymbol{V}\in\mathbb{R}^{p\times r}:\boldsymbol{V}^\top\boldsymbol{V}=\boldsymbol{I}_{r}\}$ as the collection of $p\times r$ matrices with orthonormal columns. Define the inner product of matrices $\boldsymbol{A}$ and $\boldsymbol{B}$ of the same size by $\langle\boldsymbol{A},\boldsymbol{B}\rangle=\mathrm{Tr}(\boldsymbol{A}^\top\boldsymbol{B})$. $\boldsymbol{P}_{\boldsymbol{A}}$ stands for the projection matrix onto the column space of $\boldsymbol{A}$.

\section{Problem Setup}\label{sec2}
Suppose that we observe one target dataset and $K$ independent source datasets. Specifically, for each $k\in\{0\}\cup[K]$, let $\mathcal{D}^{(k)}=\{(\boldsymbol{x}_i^{(k)},\boldsymbol{y}_i^{(k)})\}_{i=1}^{n^{(k)}}$, where $\boldsymbol{x}_i^{(k)}\in\mathbb{R}^{p}$ and $\boldsymbol{y}_i^{(k)}\in\mathbb{R}^{m}$. We refer to $\mathcal{D}^{(0)}$ as the target dataset and $\{\mathcal{D}^{(k)}\}_{k=1}^K$ as the source datasets. We assume that the observations are independent across datasets and, for each $k\in\{0\}\cup[K]$,
\begin{align*}
    (\boldsymbol{x}_i^{(k)},\boldsymbol{y}_i^{(k)})\sim\mathbb{N}_{p+m}(\boldsymbol{0}_{p+m},\boldsymbol{\Sigma}^{(k)}),\qquad\text{with}\qquad\boldsymbol{\Sigma}^{(k)}=\begin{pmatrix}
    \boldsymbol{\Sigma}_x^{(k)} & \boldsymbol{\Sigma}_{xy}^{(k)}\\
    \boldsymbol{\Sigma}_{yx}^{(k)} & \boldsymbol{\Sigma}_y^{(k)}
    \end{pmatrix}.
\end{align*}
Our primary interest lies in estimating the leading canonical directions of the target population. For the target dataset, suppose that 
\begin{align*}
    \boldsymbol{\Sigma}_{\boldsymbol{x}\boldsymbol{y}}^{(0)}=\boldsymbol{\Sigma}_{x}^{(0)}\boldsymbol{U}^{(0)}\boldsymbol{\Lambda}^{(0)}\boldsymbol{V}^{(0)\top}\boldsymbol{\Sigma}_{y}^{(0)},
\end{align*}
where $\boldsymbol{U}^{(0)}\in\mathbb{R}^{p\times r}$ and $\boldsymbol{V}^{(0)}\in\mathbb{R}^{m\times r}$ collects the $r$ canonical directions of the target population that satisfies
\begin{align*}
    \boldsymbol{U}^{(0)\top}\boldsymbol{\Sigma}_{x}^{(0)}\boldsymbol{U}^{(0)}=\boldsymbol{I}_r,\qquad \boldsymbol{V}^{(0)\top}\boldsymbol{\Sigma}_{y}^{(0)}\boldsymbol{V}^{(0)}=\boldsymbol{I}_r,
\end{align*}
and $\boldsymbol{\Lambda}^{(0)}=\mbox{diag}(\lambda_1^{(0)},\ldots,\lambda_r^{(0)})$ is a diagonal matrix. For a matrix $\boldsymbol{A}$, define its row support $\mathrm{supp}(\boldsymbol{A})$ as the indices of nonzero rows of $\boldsymbol{A}$. Our considered parameter space, denoted by $\mathcal{F}(s_u,s_v,p,m,r,\lambda;M)$, is defined as the collection of $\{\boldsymbol{\Sigma}^{(k)}\}_{k\in\{0\}\cup[K]}$:
\begin{equation}\label{eq2.1}
\left\{
    \begin{aligned}
        &|\mathrm{supp}(\boldsymbol{U}^{(0)})|\leq s_u,\quad |\mathrm{supp}(\boldsymbol{V}^{(0)})|\leq s_v,\min_{k\in[K]}\big(\lambda_{\min}(\boldsymbol{\Sigma}_{x}^{(k)})\wedge\lambda_{\min}(\boldsymbol{\Sigma}_{y}^{(k)})\big)\geq M^{-1}\\
        &\quad\max_{k\in[K]}\big(\lambda_{\max}(\boldsymbol{\Sigma}_{x}^{(k)})\vee\lambda_{\max}(\boldsymbol{\Sigma}_{y}^{(k)})\big)\leq M,~k\in\{0\}\cup[K]\\
        &\lambda_1^{(0)}\leq 1-M^{-1},\quad\lambda_r^{(0)}\geq\lambda
    \end{aligned}
    \right\}
\end{equation}
for some constant $M\geq 1$. The corresponding class of joint distributions is
\begin{equation}\label{eq2.2}
    \begin{aligned}
        \mathcal{P}:=\mathcal{P}(\boldsymbol{n},s_u,s_v,p,m,r,\lambda;M)=\Bigg\{&\bigotimes_{k=0}^{K}\left[\mathbb{N}_{p+m}(\boldsymbol{0}_{p+m},\boldsymbol{\Sigma}^{(k)})\right]^{n^{(k)}}:\\
        &\{\boldsymbol{\Sigma}^{(k)}\}_{k\in\{0\}\cup[K]}\in\mathcal{F}(s_u,s_v,p,m,r,\lambda;M)\Bigg\}
    \end{aligned},
\end{equation}
where $\boldsymbol{n}=(n^{(0)},n^{(1)},\ldots, n^{(K)})$ is the sample-size vector. To characterize the similarity between the source and target populations, we define the informative source set by
$$\mathcal{A}_h=\Big\{k\in[K]: \|\boldsymbol{\Sigma}^{(k)}-\boldsymbol{\Sigma}^{(0)}\|_2\leq h\Big\}.$$
Thus, a source dataset is regarded as informative if its joint covariance matrix is sufficiently close to that of the target population in spectral norm. 

\begin{remark}\label{remark 1}
    The global spectral-norm similarity condition $\|\boldsymbol{\Sigma}^{(k)}-\boldsymbol{\Sigma}^{(0)}\|_2\leq h$ is sufficient but stronger than necessary. It is reasonable to relax this similarity condition, e.g., assume that the following two inequalities hold:
    \begin{align*}
        &\|\boldsymbol{\Sigma}_{xy}^{(k)}-\boldsymbol{\Sigma}_{x}^{(k)}\{\boldsymbol{\Sigma}_{x}^{(0)}\}^{-1}\boldsymbol{\Sigma}_{xy}^{(0)}\|_2\vee\|\boldsymbol{\Sigma}_{yx}^{(k)}-\boldsymbol{\Sigma}_{y}^{(k)}\{\boldsymbol{\Sigma}_{y}^{(0)}\}^{-1}\boldsymbol{\Sigma}_{yx}^{(0)}\|_2\leq h,\\
        &\sup_{I,J\in[p+m]:\!|I|\leq 2s_u,|J|\leq 2s_v}\!\!\|[\boldsymbol{\Sigma}_x^{(k)}-\boldsymbol{\Sigma}_x^{(0)}]_{I,I}\|_2\!\vee\!\|[\boldsymbol{\Sigma}_{xy}^{(k)}-\boldsymbol{\Sigma}_{xy}^{(0)}]_{I,J}\|_2\!\vee\!\|[\boldsymbol{\Sigma}_y^{(k)}-\boldsymbol{\Sigma}_y^{(0)}]_{J,J}\|_2\leq h.
    \end{align*}
    The first condition controls the residuals obtained by substituting the target population regression coefficients into the source population normal equations, whereas the second controls covariance perturbations along sparse directions relevant to the target canonical subspaces. Under the same sample-size conditions as in the main theorems, the upper-bound arguments can be extended to the above similarity class. One may further introduce different informative source sets for $\boldsymbol{\Sigma}_{x},~\boldsymbol{\Sigma}_{y}$, and $\boldsymbol{\Sigma}_{xy}$. Such an extension would require block-specific pooling rules, and we leave this for future directions. Throughout this paper, we retain $\|\boldsymbol{\Sigma}^{(k)}-\boldsymbol{\Sigma}^{(0)}\|_2$ as a simple and unified similarity measure.
\end{remark}

Under the definition of an informative set, we define the oracle parameter space
\begin{equation}\label{eq2.3}
\mathcal{F}(\mathcal{A}_h):=\left\{\{\boldsymbol{\Sigma}^{(k)}\}_{k\in\{0\}\cup\mathcal{A}_h}:\qquad
    \begin{aligned}
        &\{\boldsymbol{\Sigma}^{(k)}\}_{k\in\{0\}\cup\mathcal{A}_h}\in\mathcal{F}(s_u,s_v,p,m,r,\lambda;M)\\
        &\|\boldsymbol{\Sigma}^{(k)}-\boldsymbol{\Sigma}^{(0)}\|_2\leq h,\quad\forall k\in \mathcal{A}_h
    \end{aligned}
    \right\},
\end{equation}
and the corresponding class of distributions:
\begin{equation}\label{eq2.4}
    \begin{aligned}
        \mathcal{P}(\mathcal{A}_h):=&\mathcal{P}(\boldsymbol{n},s_u,s_v,p,m,r,\lambda,h;M)\\
        =\Bigg\{&\bigotimes_{k\in\{0\}\cup\mathcal{A}_h}\left[\mathbb{N}_{p+m}(\boldsymbol{0}_{p+m},\boldsymbol{\Sigma}^{(k)})\right]^{\otimes n^{(k)}}:~\{\boldsymbol{\Sigma}^{(k)}\}_{k\in\{0\}\cup\mathcal{A}_h}\in\mathcal{F}(\mathcal{A}_h)\Bigg\}.
    \end{aligned}
\end{equation} 
We further denote the informative source sample size $n^{\mathcal{A}_h}=\sum_{k\in\mathcal{A}_h}n^{(k)}$.

To evaluate the estimation accuracy of $\boldsymbol{U}^{(0)}$, we adopt the prediction loss for $\boldsymbol{U}^{(0)}$ considered in \cite{gao2017sparse}:
\begin{align}\label{eq2.5}
    L(\hat{\boldsymbol{U}},\boldsymbol{U}^{(0)})=\inf_{\boldsymbol{W}\in\mathcal{O}(r,r)}\mathbb{E}_{\boldsymbol{x}'\mathop{=}\limits^{d}\boldsymbol{x}^{(0)}\sim\mathbb{N}(\boldsymbol{0}_p,\boldsymbol{\Sigma}_{x}^{(0)})}\|\boldsymbol{W}^\top\hat{\boldsymbol{U}}^{\top}\boldsymbol{x}'-\boldsymbol{U}^{(0)\top}\boldsymbol{x}'\|_2^2.
\end{align}
The introduction of $\boldsymbol{W}$ is to align the estimated and population canonical bases and make the loss invariant to their particular orthogonal representations. This loss function is equivalent to $$L(\hat{\boldsymbol{U}},\boldsymbol{U}^{(0)})=\inf_{\boldsymbol{W}\in\mathcal{O}(r,r)}\mathrm{Tr}\big((\hat{\boldsymbol{U}}\boldsymbol{W}-\boldsymbol{U}^{(0)})^\top\boldsymbol{\Sigma}_{x}^{(0)}(\hat{\boldsymbol{U}}\boldsymbol{W}-\boldsymbol{U}^{(0)})\big).$$
We can also define the prediction loss function for $\boldsymbol{V}^{(0)}$ symmetrically. To quantify the normalization accuracy, we define the population-level normalization error towards the target dataset as 
\begin{align}\label{eq2.6}
    \|\hat{\boldsymbol{U}}^\top\boldsymbol{\Sigma}_x^{(0)}\hat{\boldsymbol{U}}-\boldsymbol{I}_r\|_2
\end{align}
for any estimator $\hat{\boldsymbol{U}}$. Unlike exact normalization with respect to a sample covariance matrix, \eqref{eq2.6} measures the deviation of the estimated canonical coefficient vectors from the population normalization condition for the target canonical coefficients.

\section{Minimax Rates and Oracle-Trans-sCCA Algorithm}\label{sec3}
In this section, we begin by deriving minimax lower bounds that characterize the fundamental statistical limits of the problem and provide benchmarks for the achievable convergence rates. 

\begin{theorem}[Minimax Lower Bound]\label{thm: minimax lower bound}
    Suppose that $r\leq (s_u\wedge s_v)/2$. Then we have 
    \begin{align*}
        \inf_{\hat{\boldsymbol{U}}}\sup_{\mathbb{P}\in\mathcal{P}(\mathcal{A}_h)}\mathbb{P}\Bigg(\|\boldsymbol{P}_{\hat{\boldsymbol{U}}}&-\boldsymbol{P}_{\boldsymbol{U}^{(0)}}\|_{\mathrm{F}}^2\geq C_1\frac{s_u\left(r+\log(ep/s_u)\right)}{(n^{(0)}+n^{\mathcal{A}_h})\lambda^2}\\
        &+C_2\left(\frac{rh^2}{\lambda^2}\wedge \frac{s_u\left(r+\log(ep/s_u)\right)}{n^{(0)}\lambda^2}\right)\Bigg)\geq 0.8.
    \end{align*}
    for some positive constants $C_1$ and $C_2$ that depend only on $M$.
\end{theorem}

Now we introduce an oracle transfer procedure to illustrate the main ideas of our method and facilitate the subsequent theoretical analysis. Throughout this section, the informative set $\mathcal{A}_h$ and the sparsity levels $s_u$ and $s_v$ are assumed to be known. For simplicity of exposition, suppose that $n^{(k)}$ is divisible by $3$ for every $k\in\{0\}\cup\mathcal{A}_h$, and partition each dataset into three disjoint folds of equal size, denoted by $\{(\boldsymbol{x}_i^{(k)[j]}, \boldsymbol{y}_i^{(k)[j]})\}_{i\in [n^{(k)}/3]}$, $j\in[3]$. Although two folds suffice for the auxiliary datasets, we adopt a three-fold partition to align their indexing with that of the target dataset and simplify the presentation. For each $j\in[3]$, define the pooled sample covariance matrix by 
$$\hat{\boldsymbol{\Sigma}}^{[j]}=\frac{1}{(n^{(0)}+n^{\mathcal{A}_h})/3}\sum_{k\in\{0\}\cup\mathcal{A}_h}\begin{bmatrix}
    \boldsymbol{x}_i^{(k)[j]}&\boldsymbol{y}_i^{(k)[j]}
\end{bmatrix}\begin{bmatrix}
    \boldsymbol{x}_i^{(k)[j]}&\boldsymbol{y}_i^{(k)[j]}
\end{bmatrix}^\top,$$
and denote its corresponding blocks by  $\hat{\boldsymbol{\Sigma}}_x^{[j]}$, $\hat{\boldsymbol{\Sigma}}_y^{[j]}$, and $\hat{\boldsymbol{\Sigma}}_{xy}^{[j]}$. We also write $\hat{\boldsymbol{\Sigma}}^{(k)[j]}$ for the sample covariance matrix computed from the $j$-th fold of dataset $k$. 

The procedure consists of four main stages. In the first two steps, we build on the sparse CCA formulation of \cite{gao2015minimax,gao2017sparse}. The initialization step uses the pooled first-fold samples to construct an initial estimate with controlled joint loss. At this stage, we only require the initial estimation to be consistent to support the subsequent estimation steps. The transferring step then uses the pooled second-fold samples to improve estimation accuracy by incorporating information from the informative auxiliary datasets. The loss function used in this step is motivated by the regression-type reformulation of sparse CCA in \cite{gao2017sparse}. Indeed, under the canonical pair parametrization, $\boldsymbol{\Sigma}_{xy}^{(0)}\boldsymbol{V}^{(0)}=\boldsymbol{\Sigma}_{x}^{(0)}\boldsymbol{U}^{(0)}\boldsymbol{\Lambda}^{(0)}$. Hence, if $\boldsymbol{y}^{(0)\top}\boldsymbol{V}^{(0)}$ is viewed as a multivariate pseudo-response, then $\boldsymbol{U}^{(0)}\boldsymbol{\Lambda}^{(0)}$ plays the role of its population regression coefficient on $\boldsymbol{x}$. Replacing $\boldsymbol{V}^{(0)}$ by the initial estimator therefore leads naturally to the regression-type loss employed in the transferring step, together with a group-sparsity constraint to exploit the row-sparse structure.

Pooling reduces the stochastic error through the enlarged effective sample size, but the discrepancy between the source and target populations may introduce transfer bias. To correct this bias, we further consider a debiasing step using the target data. For a matrix $\boldsymbol{A}\in\mathbb{R}^{p\times r}$, define the $q$-support Frobenius norm by $\|\boldsymbol{A}\|_{\mathrm{F},q}:=\max_{S\subseteq[p],|S|\leq q}\|\boldsymbol{A}_{S,[r]}\|_{\mathrm{F}}=\sup_{\|\boldsymbol{D}\|_{\mathrm{F}}\leq 1,|\mathrm{supp}(\boldsymbol{D})|\leq q}\langle\boldsymbol{A},\boldsymbol{D}\rangle$. This quantity is a group analogue of the dual $k$-support norm; see \cite{argyriou2012sparse}. The debiasing step searches for an estimator close to the transfer estimator while enforcing a target-sample estimating-equation constraint measured by the $q$-support Frobenius norm. Finally, the normalization step converts the attained estimator into an approximately target-normalized canonical coefficient matrix. As discussed in Example \ref{example 1}, direct normalization using only the target sample covariance may introduce an error limited by target sample size, whereas direct pooled normalization may incur additional transfer bias. We therefore first normalize using the pooled covariance estimator and then apply a spectral truncation correction based on the target sample covariance. The resulting estimator allows a controlled deviation from exact target-sample normalization while preserving the optimal rate under the prediction loss. The complete procedure is summarized in Algorithm \ref{algo1}.

\begin{algorithm}[htbp]
\caption{Oracle-Trans-sCCA Algorithm}
\begin{algorithmic}[1]
\State \textbf{Input:} data $\{(\boldsymbol{x}_i^{(k)}, \boldsymbol{y}_i^{(k)})\}_{i\in [n^{(k)}], k\in \{0\}\cup[K]}$, informative set $\mathcal{A}_h$, sparsity levels $s_u,s_v$.

\State \textbf{Initialization step:} Solve the following program
\begin{align*}
    &(\hat{\boldsymbol{U}}^{[1]},\hat{\boldsymbol{V}}^{[1]})=\mathop{\arg\max}_{\boldsymbol{L}\in\mathbb{R}^{p\times r},\boldsymbol{R}\in\mathbb{R}^{m\times r}}\mathrm{Tr}(\boldsymbol{L}^\top\hat{\boldsymbol{\Sigma}}_{xy}^{[1]}\boldsymbol{R}),\\
    &\text{subject to}~\boldsymbol{L}^\top\hat{\boldsymbol{\Sigma}}_x^{[1]}\boldsymbol{L}=\boldsymbol{R}^\top\hat{\boldsymbol{\Sigma}}_y^{[1]}\boldsymbol{R}=\boldsymbol{I}_r~\text{and}~|\mathrm{supp}(\boldsymbol{L})|\leq s_u,~|\mathrm{supp}(\boldsymbol{R})|\leq s_v.
\end{align*}

\State \textbf{Transferring step:} Solve
\begin{align*}
    &\hat{\boldsymbol{U}}^{[2]}=\mathop{\arg\min}_{\boldsymbol{L}\in\mathbb{R}^{p\times r}}\mathrm{Tr}(\boldsymbol{L}^\top\hat{\boldsymbol{\Sigma}}_{x}^{[2]}\boldsymbol{L})-2\mathrm{Tr}(\boldsymbol{L}^\top\hat{\boldsymbol{\Sigma}}_{xy}^{[2]}\hat{\boldsymbol{V}}^{[1]}),~\text{subject to}~|\mathrm{supp}(\boldsymbol{L})|\leq s_u.
\end{align*}

\State \textbf{Debiasing step:} Solve
\begin{align*}
    \hat{\boldsymbol{U}}^{[2]d}=&\mathop{\arg\min}_{\boldsymbol{L}\in\mathbb{R}^{p\times r}}\left\|\boldsymbol{L}-\hat{\boldsymbol{U}}^{[2]}\right\|_{\mathrm{F}}^2,~\text{subject to}~|\mathrm{supp}(\boldsymbol{L})|\leq s_u~\text{and}~\|\hat{\boldsymbol{\Sigma}}_x^{(0)[2]}\boldsymbol{L}-\hat{\boldsymbol{\Sigma}}_{xy}^{(0)[2]}\hat{\boldsymbol{V}}^{[1]}\|_{\mathrm{F},q_u}\leq \tau_u,
\end{align*}
where $q_u=2s_u\wedge p$, and $\tau_u=C_{\tau,u}\{s_u(r+\log(ep/s_u)))/n^{(0)}\}^{1/2}$ for some constant $C_{\tau,u}>0$.

\State \textbf{Normalization step:} Let $\mathcal{H}_a(\boldsymbol{H})=\boldsymbol{Q}\mathrm{diag}
\left(\mu_\ell\mathbf{1}\{|\mu_\ell|>a\}\right)_{\ell=1}^r
\boldsymbol{Q}^\top$ be the spectral hard-thresholding operator for a symmetric matrix with eigenvalue decomposition 
$\boldsymbol{H}=\boldsymbol{Q}\mathrm{diag}(\mu_1,\ldots,\mu_r)\boldsymbol{Q}^\top$, and compute
\begin{align*}
    \hat{\boldsymbol{B}}=\hat{\boldsymbol{U}}^{[2]d\top}\hat{\boldsymbol{\Sigma}}_{x}^{[3]}\hat{\boldsymbol{U}}^{[2]d},\quad \hat{\boldsymbol{Z}}=\hat{\boldsymbol{U}}^{[2]d}\hat{\boldsymbol{B}}^{-1/2},\quad \hat{\boldsymbol{H}}=\hat{\boldsymbol{Z}}^{\top}\hat{\boldsymbol{\Sigma}}_{x}^{(0)[3]}\hat{\boldsymbol{Z}}-\boldsymbol{I}_r.
\end{align*}
The final normalized estimator is $\hat{\boldsymbol{U}}^{[3]}=\hat{\boldsymbol{Z}}\big\{\boldsymbol{I}_r+\mathcal{H}_{2\varepsilon_{0,u}}(\hat{\boldsymbol{H}})\big\}^{-1/2}$ with $\varepsilon_{0,u}=C_{\varepsilon}\big((s_u+\log(ep/s_u))/n^{(0)}\big)^{1/2}$ for some constant $C_{\varepsilon}>0$.

\State \textbf{Output:} $\hat{\boldsymbol{U}}^{[3]}$.
\end{algorithmic}
\label{algo1}
\end{algorithm}

Before analyzing the subsequent transfer and debiasing steps, we first study the statistical accuracy of the initial estimator under the joint loss. A key technical ingredient in this analysis is a refined generalized sin-theta theorem for singular subspace perturbation. The generalized sin-theta theorem of \cite{gao2015minimax} imposes a comparability condition between the leading and the smallest signal singular values, which is unnecessarily restrictive for our setting. We first establish the following refinement, which removes this condition.

\begin{theorem}[A Refined Generalized sin-theta Theorem]\label{TheoremB: Generalized sin-theta theorem without condition of same order singular values}
    Let $\boldsymbol{X},\boldsymbol{Y}\in\mathbb{R}^{p\times m}$ with singular value decomposition (SVD)
    \begin{align*}
        \boldsymbol{X}=\boldsymbol{A}_1\boldsymbol{D}_1\boldsymbol{B}_1^\top+\boldsymbol{A}_2\boldsymbol{D}_2\boldsymbol{B}_2^\top,\qquad \boldsymbol{Y}=\hat{\boldsymbol{A}}_1\hat{\boldsymbol{D}}_1\hat{\boldsymbol{B}}_1^\top+\hat{\boldsymbol{A}}_2\hat{\boldsymbol{D}}_2\hat{\boldsymbol{B}}_2^\top,
    \end{align*}
    where $\boldsymbol{D}_1=\mathrm{diag}(d_1,\ldots,d_r)$ and $\hat{\boldsymbol{D}}_1=\mathrm{diag}(\hat{d}_1,\ldots,\hat{d}_r)$ with $d_1\geq\cdots \geq d_r>0$. Suppose that there is a positive constant $\delta\in(0,d_r)$ and $\|\hat{\boldsymbol{D}}_2\|_2\leq d_r-\delta$. For either the spectral norm $\|\cdot\|_2$ or the Frobenius norm $\|\cdot\|_{\mathrm{F}}$, denoted generically by $\|\cdot\|$, define $\boldsymbol{E}=\boldsymbol{Y}-\boldsymbol{X}$, and let $\epsilon:=\|\boldsymbol{A}_1^\top\boldsymbol{E}\|\vee\|\boldsymbol{E}\boldsymbol{B}_1\|$ be the error under the norm $\|\cdot\|$. Then we have
    \begin{align*}
        \|\boldsymbol{A}_1\boldsymbol{B}_1^\top-\hat{\boldsymbol{A}}_1\hat{\boldsymbol{B}}_1^\top\|\leq\frac{2\epsilon}{\delta},
    \end{align*}
    and 
    \begin{align*}
        \|\boldsymbol{A}_1\boldsymbol{D}_1\boldsymbol{B}_1^\top-\hat{\boldsymbol{A}}_1\hat{\boldsymbol{D}}_1\hat{\boldsymbol{B}}_1^\top\|\leq\Bigg(\frac{\sqrt{2}(d_1+\hat{d}_1)}{\delta}+1\Bigg)\epsilon.
    \end{align*}
\end{theorem}

The proof of Theorem \ref{TheoremB: Generalized sin-theta theorem without condition of same order singular values} is deferred to Section B.1 of the Supplementary Material.

\begin{remark}
    Theorem \ref{TheoremB: Generalized sin-theta theorem without condition of same order singular values} extends Theorem 5 of \cite{gao2015minimax} by removing the condition $d_1\vee \hat{d}_1\leq C d_r$ imposed in Theorem 5 of \cite{gao2015minimax}. The key idea is to reformulate the singular-subspace perturbation problem through the associated Sylvester equation, which allows the perturbation to be controlled by the relevant spectral gap without requiring the leading signal singular values to be of the same order. In the CCA setting, this in particular removes the requirement $\lambda_1^{(0)}\leq C\lambda_r^{(0)}$ used in \cite{gao2015minimax}. This refinement enables us to analyze the initial estimator without imposing comparability among the nonzero canonical correlations. Beyond the present application, this result may also be useful in other singular-subspace perturbation problems where the leading signal singular values are not comparable to the smallest one and may be of interest in its own right.
\end{remark}

Equipped with Theorem \ref{TheoremB: Generalized sin-theta theorem without condition of same order singular values}, we now characterize the estimation error of the initial estimator under the joint loss. As a by-product of this analysis, we also establish the minimax lower bound under the same loss, which allows us to compare the accuracy of the initialization with the fundamental statistical limit.

\begin{theorem}[Initial Estimation Error and Minimax Lower Bound under Joint Loss]\label{thm: initial estimation error and minimax lower bound for joint loss}
    Define the rate $$\mathcal{C}_{u,v}=r(s_u+s_v)+s_u\log(ep/s_u)+s_v\log(em/s_v).$$ We have 
    \begin{align*}
        \inf_{\hat{\boldsymbol{U}},\hat{\boldsymbol{V}}}\sup_{\mathbb{P}\in\mathcal{P}(\mathcal{A}_h)}\mathbb{P}\Bigg(\Big\|\hat{\boldsymbol{U}}\hat{\boldsymbol{V}}^\top-\boldsymbol{U}^{(0)}&\boldsymbol{V}^{(0)\top}\Big\|_{\mathrm{F}}^2\geq C\frac{\mathcal{C}_{u,v}}{(n^{(0)}+n^{\mathcal{A}_h})\lambda^2}\\
        &~~~~~~~~~~~~~~~~~+C'\left(\frac{rh^2}{\lambda^2}\wedge \frac{\mathcal{C}_{u,v}}{n^{(0)}\lambda^2}\right)\Bigg)\geq 0.8
    \end{align*}
    for some positive constants $C$ and $C'$ that depend only on $M$. Moreover, assume that $$\sqrt{\frac{s_u+\log(ep/s_u)+s_v+\log(em/s_v)}{n^{\mathcal{A}_h}+n^{(0)}}}+h\frac{n^{\mathcal{A}_h}}{n^{\mathcal{A}_h}+n^{(0)}}\leq c_1\lambda,$$
    and $\mathcal{C}_{u,v}\leq c_2N_h$ for some sufficiently small constants $c_1,~c_2>0$. Then there exist constants $C,C'>0$ such that
    \begin{align*}
        &\left\|\hat{\boldsymbol{U}}^{[1]}\hat{\boldsymbol{V}}^{[1]\top}-\boldsymbol{U}^{(0)}\boldsymbol{V}^{(0)\top}\right\|_{\mathrm{F}}^2\leq C\left\{\frac{\mathcal{C}_{u,v}}{(n^{\mathcal{A}_h}+n^{(0)})\lambda^2}+\frac{rh^2}{\lambda^2}\frac{(n^{\mathcal{A}_h})^2}{(n^{\mathcal{A}_h}+n^{(0)})^2}\right\}
    \end{align*}
    with probability at least $1-\exp\left\{-C'\left(s_u+\log(ep/s_u)\right)\right\}-\exp\left\{-C'\left(s_v+\log(em/s_v)\right)\right\}$.
\end{theorem}

It is worth emphasizing that the initial estimator is generally not minimax optimal under the joint loss. This sub-optimality has two sources. Even in the single-dataset setting, the corresponding initial estimator of \cite{gao2017sparse} is only required to be consistent and is not itself minimax optimal. In our transfer setting, source-target heterogeneity introduces an additional bias. The subsequent transferring and debiasing steps jointly refine the initial estimator: the transfer step exploits information from informative sources, whereas the debiasing step corrects the bias induced by source-target heterogeneity. Since only consistency is needed at the initialization stage, we do not further refine the initial estimator before the debiasing step.

We next show that, after the transferring, debiasing, and truncated normalization steps, the oracle estimator produced by Algorithm \ref{algo1} attains the minimax optimal rate under the prediction loss.

\begin{theorem}[Bound for Oracle Estimator]\label{thm: bound for oracle estimator}
    Under the conditions of Theorem \ref{thm: initial estimation error and minimax lower bound for joint loss}, further assume that $\left(s_u\log(ep/s_u)/n^{(0)}\right)\vee \left(hn^{\mathcal{A}_h}/(n^{\mathcal{A}_h}+n^{(0)})\right)\leq c$ for some sufficiently small constant $c$, then  there exist constants $C,C'>0$ such that $\hat{\boldsymbol{U}}^{[3]}$ produced by Algorithm \ref{algo1} satisfies
    \begin{align*}
        &L(\hat{\boldsymbol{U}}^{[3]},\boldsymbol{U}^{(0)})\leq C\left\{\frac{rs_u+s_u\log(ep/s_u)}{(n^{\mathcal{A}_h}+n^{(0)})\lambda^2}+\frac{rh^2}{\lambda^2}\frac{(n^{\mathcal{A}_h})^2}{(n^{\mathcal{A}_h}+n^{(0)})^2}\wedge\frac{rs_u+s_u\log(ep/s_u)}{n^{(0)}\lambda^2}\right\},
    \end{align*}
    with probability at least $1-\exp\left\{-C'\left(s_u+\log(ep/s_u)\right)\right\}-\exp\left\{-C'\left(s_v+\log(em/s_v)\right)\right\}$. Moreover, with the same probability, we have the population-level normalization error bound:
    \begin{align*}
        \|\hat{\boldsymbol{U}}^{[3]\top}\!\boldsymbol{\Sigma}_{x}^{(0)}\hat{\boldsymbol{U}}^{[3]}\!\!-\!\boldsymbol{I}_r\|_2\!\leq \!C''\!\!\left(\!\sqrt{\frac{s_u+\log(ep/s_u)}{n^{\mathcal{A}_h}+n^{(0)}}}\!+\!\left(\!h\frac{n^{\mathcal{A}_h}}{n^{\mathcal{A}_h}+n^{(0)}}\!\wedge\!\sqrt{\frac{s_u+\log(ep/s_u)}{n^{(0)}}}\right)\!\right).
    \end{align*}
\end{theorem}

\begin{remark}\label{remark: normalization comparison}
    It can be shown that the population-level normalization error \eqref{eq2.6} of the estimator in \cite{gao2017sparse} is of the order $\sqrt{(s_u+\log(ep/s_u))/n^{(0)}}$ with high probability by applying Lemma B.5 in the Supplementary Material. This shows that our population-level normalization error is also smaller.
\end{remark}

\section{Adaptive and Computationally Feasible Estimation}\label{sec4}

\subsection{Adaptive and Computationally Feasible Estimation with Known Informative Set}\label{sec4.1}

Algorithm \ref{algo1} serves as an oracle benchmark because the informative source set $\mathcal{A}_h$ and the sparsity levels $s_u,~s_v$ are all assumed to be known, and the algorithm is not computationally efficient, as it requires traversing all subsets with restricted support size. We first retain the assumption that the informative set $\mathcal{A}_h$ is known and focus on developing a computationally feasible procedure. The practical implementation of source detection is deferred to Section \ref{sec4.2}.

For the initialization step in Algorithm \ref{algo1}, we replace the sparsity-constrained optimization by a convex relaxation. Specifically, letting $\boldsymbol{F}$ serve as a surrogate of $\boldsymbol{L}\boldsymbol{R}^\top$, we consider the penalized objective 
\begin{align*}
    \max_{\boldsymbol{F}\in\mathbb{R}^{p\times m}}\langle\hat{\boldsymbol{\Sigma}}_{xy}^{[1]},\boldsymbol{F}\rangle-\eta\|\boldsymbol{F}\|_1,
\end{align*}
where $\eta$ is a penalty parameter and $\|\boldsymbol{F}\|_1=\sum_{i\in[p],j\in[m]}|F_{ij}|$ is the element-wise $\ell_1$ norm. By construction, we have
$$\{\hat{\boldsymbol{\Sigma}}_x^{[1]}\}^{1/2}\boldsymbol{F}\{\hat{\boldsymbol{\Sigma}}_y^{[1]}\}^{1/2}\in\mathcal{O}_r=\{\boldsymbol{A}\boldsymbol{B}^\top:\boldsymbol{A}\in\mathcal{O}(p,r),\boldsymbol{B}\in\mathcal{O}(m,r)\}$$
Following \cite{gao2017sparse}, we relax $\mathcal{O}_r$ to its convex hull $$\mathcal{C}_r:=\{\boldsymbol{A}\in\mathbb{R}^{p\times m}:\|\boldsymbol{A}\|_*\leq r,~\|\boldsymbol{A}\|_2\leq 1\},$$ which leads to the initialization step in Algorithm \ref{algo2}. 

Let $\hat{\boldsymbol{F}}^{[1]}$ be the initial estimator, and $(\hat{\boldsymbol{U}}^{[1]},\hat{\boldsymbol{V}}^{[1]})$ be the first $r$ left and right singular vectors of $\hat{\boldsymbol{F}}^{[1]}$. In the transferring step, we remove the sparsity constraints and consider a group-Lasso penalization \citep{yuan2006model}. Since the resulting estimator is no longer exactly sparse, the debiasing step in Algorithm \ref{algo1} is no longer directly applicable. We therefore adopt an aggregation scheme that adaptively selects between the transfer estimator and the target-only estimator, in the spirit of \cite{li2022transfer,li2024estimation}. Finally, we apply the same truncated normalization procedure as in Algorithm \ref{algo1}.

\begin{algorithm}
\caption{Trans-sCCA Algorithm}
\begin{algorithmic}[1]
\State \textbf{Input:} data $\{(\boldsymbol{x}_i^{(k)}, \boldsymbol{y}_i^{(k)})\}_{i\in [n^{(k)}], k\in \{0\}\cup[K]}$, informative set $\mathcal{A}_h$, sparsity levels $s_u,s_v$.

\State \textbf{Initialization step:} Let $\eta\asymp (\log(p+m)/N_h)^{1/2}+hn^{\mathcal{A}_h}/N_h$. Solve the following program
\begin{align*}
    &\hat{\boldsymbol{F}}^{[1]}=\max_{\boldsymbol{F}\in\mathbb{R}^{p\times m}}\langle\hat{\boldsymbol{\Sigma}}_{xy}^{[1]},\boldsymbol{F}\rangle-\eta\|\boldsymbol{F}\|_1,\\
    &\text{subject to}~\left\|\{\hat{\boldsymbol{\Sigma}}_x^{[1]}\}^{1/2}\boldsymbol{F}\{\hat{\boldsymbol{\Sigma}}_y^{[1]}\}^{1/2}\right\|_*\leq r,~\text{and}~\left\|\{\hat{\boldsymbol{\Sigma}}_x^{[1]}\}^{1/2}\boldsymbol{F}\{\hat{\boldsymbol{\Sigma}}_y^{[1]}\}^{1/2}\right\|_2\leq 1,
\end{align*}
and define $(\hat{\boldsymbol{U}}^{[1]},\hat{\boldsymbol{V}}^{[1]})$ to be the first $r$ left and right singular vectors of $\hat{\boldsymbol{F}}^{[1]}$.

\State \textbf{Transferring step:} Solve $\hat{\boldsymbol{U}}^{[2]}=\mathop{\arg\min}_{\boldsymbol{L}\in\mathbb{R}^{p\times r}}\mathrm{Tr}(\boldsymbol{L}^\top\hat{\boldsymbol{\Sigma}}_{x}^{[2]}\boldsymbol{L})-2\mathrm{Tr}(\boldsymbol{L}^\top\hat{\boldsymbol{\Sigma}}_{xy}^{[2]}\hat{\boldsymbol{V}}^{[1]})+\eta_{t,u}\|\boldsymbol{L}\|_{2,1}.$

\State \textbf{Aggregation step:} Solve 
\begin{align*}
    &\hat{\boldsymbol{U}}^{(0)[2]}=\mathop{\arg\min}_{\boldsymbol{L}\in\mathbb{R}^{p\times r}}\mathrm{Tr}(\boldsymbol{L}^\top\hat{\boldsymbol{\Sigma}}_{x}^{(0)[2]}\boldsymbol{L})-2\mathrm{Tr}(\boldsymbol{L}^\top\hat{\boldsymbol{\Sigma}}_{xy}^{(0)[2]}\hat{\boldsymbol{V}}^{[1]})+\eta_{0,u}\|\boldsymbol{L}\|_{2,1}.
\end{align*}
Define $\hat{Q}_0^{[3]}(\boldsymbol{L})=\mathrm{Tr}(\boldsymbol{L}^\top\hat{\boldsymbol{\Sigma}}_{x}^{(0)[3]}\boldsymbol{L})-2\mathrm{Tr}(\boldsymbol{L}^\top\hat{\boldsymbol{\Sigma}}_{xy}^{(0)[3]}\hat{\boldsymbol{V}}^{[1]})$, and let $$\hat{\boldsymbol{U}}_{A}^{[2]}=\begin{cases}
    \hat{\boldsymbol{U}}^{[2]},&~\text{if}~\hat{Q}_0^{[3]}(\hat{\boldsymbol{U}}^{[2]})\leq \hat{Q}_0^{[3]}(\hat{\boldsymbol{U}}^{(0)[2]})+\varrho_u,\\
    \hat{\boldsymbol{U}}^{(0)[2]},&~\text{otherwise}.
\end{cases}$$

\State \textbf{Normalization step:} Let $\mathcal{H}_a(\boldsymbol{H})=\boldsymbol{Q}\mathrm{diag}
\left(\mu_\ell\mathbf{1}\{|\mu_\ell|>a\}\right)_{\ell=1}^r
\boldsymbol{Q}^\top$ be the spectral hard-thresholding operator for a symmetric matrix with eigenvalue decomposition 
$\boldsymbol{H}=\boldsymbol{Q}\mathrm{diag}(\mu_1,\ldots,\mu_r)\boldsymbol{Q}^\top$, and compute
\begin{align*}
    \hat{\boldsymbol{B}}_A=\hat{\boldsymbol{U}}_A^{[2]\top}\hat{\boldsymbol{\Sigma}}_{x}^{[3]}\hat{\boldsymbol{U}}_A^{[2]},\quad \hat{\boldsymbol{Z}}_A=\hat{\boldsymbol{U}}^{[2]}_A\hat{\boldsymbol{B}}_A^{-1/2},\quad \hat{\boldsymbol{H}}_A=\hat{\boldsymbol{Z}}_A^{\top}\hat{\boldsymbol{\Sigma}}_{x}^{(0)[3]}\hat{\boldsymbol{Z}}_A-\boldsymbol{I}_r.
\end{align*}
The final normalized estimator is $\hat{\boldsymbol{U}}^{[3]}=\hat{\boldsymbol{Z}}_A\big\{\boldsymbol{I}_r+\mathcal{H}_{2\varepsilon_{0,u}}(\hat{\boldsymbol{H}}_A)\big\}^{-1/2}$.

\State \textbf{Output:} $\hat{\boldsymbol{U}}^{[3]}$.
\end{algorithmic}
\label{algo2}
\end{algorithm}

\begin{theorem}\label{thm: Computaional efficient algorithm}
    Assume that $s_us_v\log(p+m)+hn^{\mathcal{A}_h}(n^{\mathcal{A}_h}+n^{(0)})\leq c(n^{\mathcal{A}_h}+n^{(0)})$ for some sufficiently small constants $c>0$ and $$\left(s_us_v\log(p+m)/(n^{\mathcal{A}_h}+n^{(0)})+s_us_vh^2(n^{\mathcal{A}_h}/(n^{\mathcal{A}_h}+n^{(0)}))^2\right)/\lambda^2\leq C$$ for some constant $C>0$. Choosing $\eta\asymp (\log(p+m)/(n^{\mathcal{A}_h}+n^{(0)}))^{1/2}+hn^{\mathcal{A}_h}/(n^{\mathcal{A}_h}+n^{(0)})$, $\eta_{t,u}\asymp ((r+\log (ep/s_u))/(n^{\mathcal{A}_h}+n^{(0)}))^{1/2}$, $\eta_{0,u}\asymp ((r+\log (ep/s_u))/n^{(0)})^{1/2}$, $\varrho_u\asymp (s_ur+s_u\log (ep/s_u))/n^{(0)}$, and $\varepsilon_{0,u}\asymp\big((s_u+\log(ep/s_u))/n^{(0)}\big)^{1/2}$, there exist constants $C,C'>0$ such that
    \begin{align*}
        &L(\hat{\boldsymbol{U}}^{[3]},\boldsymbol{U}^{(0)})\leq C\left\{\frac{rs_u+s_u\log(ep/s_u)}{(n^{\mathcal{A}_h}+n^{(0)})\lambda^2}+\frac{rh^2}{\lambda^2}\frac{(n^{\mathcal{A}_h})^2}{(n^{\mathcal{A}_h}+n^{(0)})^2}\wedge\frac{rs_u+s_u\log(ep/s_u)}{n^{(0)}\lambda^2}\right\},
    \end{align*}
    with probability at least $1-\exp(-C'(s_u+\log(ep/s_u)))-\exp(-C'(s_v+\log(em/s_v)))$. Moreover, with the same probability, we have the population-level normalization error bound:
    \begin{align*}
        \|\hat{\boldsymbol{U}}^{[3]\top}\!\boldsymbol{\Sigma}_{x}^{(0)}\hat{\boldsymbol{U}}^{[3]}\!\!-\!\boldsymbol{I}_r\|_2\!\leq \!C''\!\!\left(\!\sqrt{\frac{s_u+\log(ep/s_u)}{n^{\mathcal{A}_h}+n^{(0)}}}\!+\!\left(\!h\frac{n^{\mathcal{A}_h}}{n^{\mathcal{A}_h}+n^{(0)}}\!\wedge\!\sqrt{\frac{s_u+\log(ep/s_u)}{n^{(0)}}}\right)\!\right).
    \end{align*}
\end{theorem}

\begin{remark}\label{remark 2}
    We note that the additional requirements on the sample size and the similarity level $h$ arise from the initialization step, for which a consistent initial estimator is needed. Alternatively, the initialization can be constructed using only the target dataset. In this case, we may choose $\eta\asymp(\log(p+m)/n^{(0)})^{1/2}$, and it suffices to require $s_us_v\log(p+m)\leq cn^{(0)}$ for some sufficiently small constants $c>0$, $s_us_v\log(p+m)/(n^{(0)}\lambda^2)\leq C$ for some constant $C>0$, and $r/(r+\log (ep/s_u))\cdot h^2(n^{\mathcal{A}_h}/(n^{\mathcal{A}_h}+n^{(0)}))^2\leq c'$ for some sufficiently small constant $c'$. This substantially relaxes the restriction on the similarity level $h$, at the cost of requiring the target sample alone to be sufficient for consistent initialization. 
\end{remark}

\subsection{Identifying a Potential Informative Set for aggregation}\label{sec4.2}
Algorithm \ref{algo2} assumes that the informative source set $\mathcal{A}_h$ is known and performs an aggregation step to avoid negative transfer. As discussed in \ref{remark 2}, when $h\leq c$ for some sufficiently small constant $c>0$, the Trans-sparse CCA estimator with target-only initialization and the subsequent aggregation step remains minimax optimal. Nevertheless, including some non-informative auxiliary datasets that have large heterogeneity may still weaken the improvement of transfer learning. This motivates a data-driven procedure for constructing a candidate informative set before applying Algorithm \ref{algo2}.

The basic idea is to assess each source dataset according to whether augmenting the target training sample with this source improves the canonical correlation on held-out target observations. We first rank the source datasets according to their potential contribution to the target CCA problem. A similar ranking strategy has also been applied or discussed in \cite{li2022transfer,li2023transfer,li2024estimation,zhao2026trans}. It is pointed out in \cite{hanneke2022no} that if the rank of the auxiliary research can be recovered, then empirical risk minimization via cross-fitting can achieve adaptation to the information set to some extent. 

Let $\{\mathcal{I}_1,\ldots,\mathcal{I}_J\}$ be a $J$-fold partition of the target observations, and write $\mathcal{I}_{-\ell}=[n^{(0)}]\setminus \mathcal{I}_{\ell}$ for each $\ell\in[J]$. For each $\ell$, let $\hat{\boldsymbol{\Sigma}}^{(0)[-\ell]}$ denote the sample covariance matrix computed from the target observations indexed by $\mathcal{I}_{-\ell}$. For source $k$, define the pooled covariance matrix 
$$\hat{\boldsymbol\Sigma}_{k}^{[-\ell]}=\frac{|\mathcal{I}_{-\ell}|\hat{\boldsymbol{\Sigma}}^{(0)[-\ell]}+n^{(k)}\hat{\boldsymbol{\Sigma}}^{(k)}
}{|\mathcal{I}_{-\ell}|+n^{(k)}},$$
with corresponding blocks $\hat{\boldsymbol\Sigma}_{k,x}^{[-\ell]}$, $\hat{\boldsymbol\Sigma}_{k,y}^{[-\ell]}$, $\hat{\boldsymbol\Sigma}_{k,xy}^{[-\ell]}$, and $\hat{\boldsymbol\Sigma}_{k,yx}^{[-\ell]}$. For each validation fold $\ell\in[J]$, let $(\hat{\boldsymbol{U}}_{\mathrm{pil}}^{[-\ell]},\hat{\boldsymbol{V}}_{\mathrm{pil}}^{[-\ell]})$ be attained via initialization step constructed using only the target observations indexed by $\mathcal{I}_{-\ell}$. Using the same training sample, we then form the source-augmented estimators and the estimators using only the target training sample:
\begin{align*}
    &\hat{\boldsymbol{U}}_{k}^{[-\ell]}=\mathop{\arg\min}_{\boldsymbol{L}\in\mathbb{R}^{p\times r}}\Bigg\{\mathrm{Tr}\left(\boldsymbol{L}^\top\hat{\boldsymbol{\Sigma}}_{x,k}^{[-\ell]}\boldsymbol{L}\right)-2\mathrm{Tr}\left(\boldsymbol{L}^\top\hat{\boldsymbol{\Sigma}}_{xy,k}^{[-\ell]}\hat{\boldsymbol{V}}_{\mathrm{pil}}^{[-\ell]}\right)+\eta_{u,k}\|\boldsymbol{L}\|_{2,1}\Bigg\},\\
    &\hat{\boldsymbol{V}}_{k}^{[-\ell]}=\mathop{\arg\min}_{\boldsymbol{R}\in\mathbb{R}^{m\times r}}\Bigg\{\mathrm{Tr}\left(\boldsymbol{R}^\top\hat{\boldsymbol{\Sigma}}_{y,k}^{[-\ell]}\boldsymbol{R}\right)-2\mathrm{Tr}\left(\boldsymbol{R}^\top\hat{\boldsymbol{\Sigma}}_{yx,k}^{[-\ell]}\hat{\boldsymbol{U}}_{\mathrm{pil}}^{[-\ell]}\right)+\eta_{v,k}\|\boldsymbol{R}\|_{2,1}\Bigg\},\\
    &\hat{\boldsymbol{U}}_{0}^{[-\ell]}=\mathop{\arg\min}_{\boldsymbol{L}\in\mathbb{R}^{p\times r}}\Bigg\{\mathrm{Tr}\left(\boldsymbol{L}^\top\hat{\boldsymbol{\Sigma}}_{x}^{(0)[-\ell]}\boldsymbol{L}\right)-2\mathrm{Tr}\left(\boldsymbol{L}^\top\hat{\boldsymbol{\Sigma}}_{xy}^{(0)[-\ell]}\hat{\boldsymbol{V}}_{\mathrm{pil}}^{[-\ell]}\right)+\eta_{u,0}\|\boldsymbol{L}\|_{2,1}\Bigg\},\\
    &\hat{\boldsymbol{V}}_{0}^{[-\ell]}=\mathop{\arg\min}_{\boldsymbol{R}\in\mathbb{R}^{m\times r}}\Bigg\{\mathrm{Tr}\left(\boldsymbol{R}^\top\hat{\boldsymbol{\Sigma}}_{y}^{(0)[-\ell]}\boldsymbol{R}\right)-2\mathrm{Tr}\left(\boldsymbol{R}^\top\hat{\boldsymbol{\Sigma}}_{yx}^{(0)[-\ell]}\hat{\boldsymbol{U}}_{\mathrm{pil}}^{[-\ell]}\right)+\eta_{v,0}\|\boldsymbol{R}\|_{2,1}\Bigg\}.
\end{align*}
On the $\ell$-th fold, let $\widehat{\mathrm{CC}}_{0}(\ell)$ and $\widehat{\mathrm{CC}}_{k}(\ell)$ denote the sums of the $r$ canonical correlations evaluated on it for each $k\in[K]$. We define the empirical transfer gain of source $k$ as $\widehat{\mathrm{CC}}_{k}(\ell)-\widehat{\mathrm{CC}}_{0}(\ell)$, where a larger value indicates greater potential benefit from transferring source $k$. The sources are then ranked in decreasing order of $\sum_{\ell\in[J]}\big(\widehat{\mathrm{CC}}_{k}(\ell)-\widehat{\mathrm{CC}}_{0}(\ell)\big)/J$. Let $k_{(1)},\ldots,k_{(K)}$ denote the resulting ordering and define the nested candidate sets and $\widehat{\mathcal{A}}_q=\{k_{(1)},\ldots,k_{(q)}\}$ for $q\in[K]$ with $\widehat{\mathcal{A}}_0=\emptyset$. To determine the number of sources to retain, we use a second $J$-fold partition of the target observations. For each $q\in\{0\}\cup[K]$ and each validation fold, the target-only initializer and the transfer estimator based on the sources in $\widehat{\mathcal{A}}_q$ are fitted using the complementary target training observations and are evaluated on the held-out target fold. We then choose $\hat{q}=\mathop{\arg\max}_{q\in\{0\}\cup[K]} \widehat{\mathrm{CC}}(q)$, where $\widehat{\mathrm{CC}}(q)$ is the resulting cross-validated sum of the first $r$ canonical correlations. The estimated informative set is finally given by $\widehat{\mathcal{A}}_{\hat{q}}$. Note that we use the same training observations for initialization and subsequent estimation here to avoid further reducing the already limited target sample size in practice.

\section{Numerical Experiments}\label{sec5}
In this section, we provide numerical experiments to illustrate the finite-sample performance of our proposed Trans-sCCA estimator.

\subsection{Simulation Results}\label{sec5.1}

We consider the setting where $p=m$, $n^{(0)}=n^{(k)}=n$ for all $k\in[K]$, and $K=10$. We take $$(p,m,n)\in\{(150,150,200),(150,150,400),(300,300,200),(300,300,400)\}.$$ For the target population, we set $\boldsymbol{\Sigma}_x^{(0)}=\boldsymbol{\Sigma}_y^{(0)}=(\sigma_{ij}^{(0)})$, where $\sigma_{ij}^{(0)}=0.3^{|i-j|}$. The row supports of both $\boldsymbol{U}^{(0)}$ and $\boldsymbol{V}^{(0)}$ are fixed as $S=\{1,3,5,\ldots,19\}$ with $|S|=s=10$. To generate the leading canonical directions, we first generate matrices $\boldsymbol{A}_u$ and $\boldsymbol{A}_v$, whose entries on the support rows are independently sampled from the uniform distribution on the set $\{-2, -1, 0, 1, 2\}$, and whose remaining rows are zero. We then set $\boldsymbol{U}^{(0)}=\boldsymbol{A}_u(\boldsymbol{A}_u^\top\boldsymbol{\Sigma}_x^{(0)}\boldsymbol{A}_u)^{-1/2}$ and $\boldsymbol{V}^{(0)}=\boldsymbol{A}_v(\boldsymbol{A}_v^\top\boldsymbol{\Sigma}_y^{(0)}\boldsymbol{A}_v)^{-1/2}$ such that $\boldsymbol{U}^{(0)}$ and $\boldsymbol{V}^{(0)}$ are normalized with respect to $\boldsymbol{\Sigma}_x^{(0)}$ and $\boldsymbol{\Sigma}_y^{(0)}$, respectively. We consider both exactly and approximately low-rank settings. In the exactly low-rank setting, $r=2$, $\lambda_1^{(0)}=0.9$, and $\lambda_2^{(0)}=0.8$, and $\boldsymbol{\Sigma}_{xy}^{(0)}=\boldsymbol{\Sigma}_x^{(0)}\boldsymbol{U}^{(0)}\boldsymbol{\Lambda}^{(0)}\boldsymbol{V}^{(0)\top}\boldsymbol{\Sigma}_y^{(0)}$. In the approximately low-rank setting, we augment $\boldsymbol{U}^{(0)}$ and $\boldsymbol{V}^{(0)}$ with additional columns to form matrices $\boldsymbol{U}$ and $\boldsymbol{V}$ satisfying $\boldsymbol{U}^{\top}\boldsymbol{\Sigma}_x^{(0)}\boldsymbol{U}=\boldsymbol{V}^{\top}\boldsymbol{\Sigma}_y^{(0)}\boldsymbol{V}=\boldsymbol{I}_p$, and set $\lambda_1^{(0)}=0.9$, $\lambda_2^{(0)}=0.8$, $\lambda_i^{(0)}=0.8 e^{-2(i-2)}$ for $i\in\{3,\ldots,p\}$. We still set $r=2$ in the simulation implementations in this case. We then define $\boldsymbol{\Sigma}_{xy}^{(0)}=\boldsymbol{\Sigma}_x^{(0)}\boldsymbol{U}\boldsymbol{\Lambda}^{(0)}\boldsymbol{V}^{\top}\boldsymbol{\Sigma}_y^{(0)}$. Denote the resulting joint target covariance matrix by $\boldsymbol{\Sigma}^{(0)}$. 

For each source $k$, we first construct an auxiliary endpoint covariance matrix. We independently generate $\rho_x^{(k)}$ and $\rho_y^{(k)}$ from the uniform distribution on the set $[0.1,0.2]\cup[0.4,0.6]$, and set $[\tilde{\boldsymbol{\Sigma}}_x^{(k)}]_{i,j}=(\rho_x^{(k)})^{|i-j|},[\tilde{\boldsymbol{\Sigma}}_y^{(k)}]_{i,j}=(\rho_y^{(k)})^{|i-j|}$. The common row support $S^{(k)}$ of $\tilde{\boldsymbol{U}}^{(k)}$ and $\tilde{\boldsymbol{V}}^{(k)}$ is randomly chosen from $\{3s+1,\ldots,p\}$ with $|S^{(k)}|=s=10$. The corresponding loading matrices $\tilde{\boldsymbol{U}}^{(k)}$ and $\tilde{\boldsymbol{V}}^{(k)}$ are generated using the same scheme as for the target and normalized with respect to $\tilde{\boldsymbol{\Sigma}}_x^{(k)}$ and $\tilde{\boldsymbol{\Sigma}}_y^{(k)}$, respectively. We independently generate $\tilde{\lambda}_1^{(k)}\sim\mathrm{Unif}(0.85,0.95)$ and $\tilde{\lambda}_2^{(k)}\sim\mathrm{Unif}(0.65,0.85)$. In the exactly low-rank setting, these are the only two nonzero canonical correlations. In the approximately low-rank setting, we augment $\tilde{\boldsymbol{U}}^{(k)}$ and $\tilde{\boldsymbol{V}}^{(k)}$ using the same scheme as for the target, and additionally generate $c_k\sim \mathrm{Unif}(0.5,2)$ and set $\tilde{\lambda}_i^{(k)}=\tilde{\lambda}_2^{(k)} e^{-c_k(i-2)}$ for $i\in\{3,\ldots,p\}$. The resulting source covariance matrix is denoted by $\tilde{\boldsymbol{\Sigma}}^{(k)}$. We then let $\boldsymbol{\Sigma}^{(k)}=\boldsymbol{\Sigma}^{(0)}+0.05(\tilde{\boldsymbol{\Sigma}}^{(k)}-\boldsymbol{\Sigma}^{(0)})/\|\tilde{\boldsymbol{\Sigma}}^{(k)}-\boldsymbol{\Sigma}^{(0)}\|_2$ for informative sets and $\boldsymbol{\Sigma}^{(k)}=\tilde{\boldsymbol{\Sigma}}^{(k)}$ for non-informative sets. We let the number of the informative set $\mathcal{A}_h$, denoted by $K_A$, range from $0$ to $K$, and the informative sources are selected completely at random. Finally, observations from both the target and source populations are independently generated from centered multivariate Gaussian distributions with their corresponding joint covariance matrices.

We first compare three estimators, including the Trans-sCCA with known informative set (denoted by \texttt{Oracle-Trans-sCCA} in simulations for notational simplicity, Algorithm \ref{algo2} with $\mathcal{A}_h$ as input), the \texttt{Pooled-sCCA} (Run initial and transferring steps in Algorithm \ref{algo2} with all datasets as input and apply a target normalization), and the \texttt{Target-sCCA} (Algorithm \ref{algo2} with only target dataset). For the initialization step, we adopt the ADMM-based optimization procedure in Section 10 of \citet{gao2017sparse}, which solves the convex relaxation of the canonical correlation initialization problem. Since the initialization involves repeated singular value decompositions in the ADMM iterations, numerical failures of the SVD routine may occasionally occur in high-dimensional settings, particularly for \texttt{Target-sCCA}. To ensure a fair comparison among competing methods, we summarize the empirical results based only on the replications for which the compared methods successfully produce valid estimates. For the penalty parameters in the initialization step, we use the same penalty $0.55\sqrt{\log (p+m)/n_{\mathrm{eff}}}$, where $n_{\mathrm{eff}}$ is the effective sample size for each method (e.g., $n_{\mathrm{eff}}=n(K_A+1)$ for \texttt{Oracle-Trans-sCCA}, $n_{\mathrm{eff}}=n(K+1)$ for \texttt{Pooled-sCCA}, and $n_{\mathrm{eff}}=n$ for \texttt{Target-sCCA}). All methods select the group-Lasso penalty parameters ($\eta_{t,u}=b\sqrt{(r+\log p)/n_{\mathrm{eff}}}$ and $\eta_{0,u}=b\sqrt{(r+\log p)/n}$, and symmetrically, $\eta_{t,v}=b\sqrt{(r+\log m)/n_{\mathrm{eff}}}$ and $\eta_{0,v}=b\sqrt{(r+\log m)/n}$ for $\boldsymbol{V}^{(0)}$-side) in the transferring step or the aggregation step by five-fold cross-validation for the target dataset, where the criterion is to maximize the sums of the $r$ canonical correlations from the candidate set $b\in\{0.5,1,1.5,2\}$ similar to Section \ref{sec4.2}.

\begin{figure}[!ht]
    \centering
    \includegraphics[width=\textwidth]{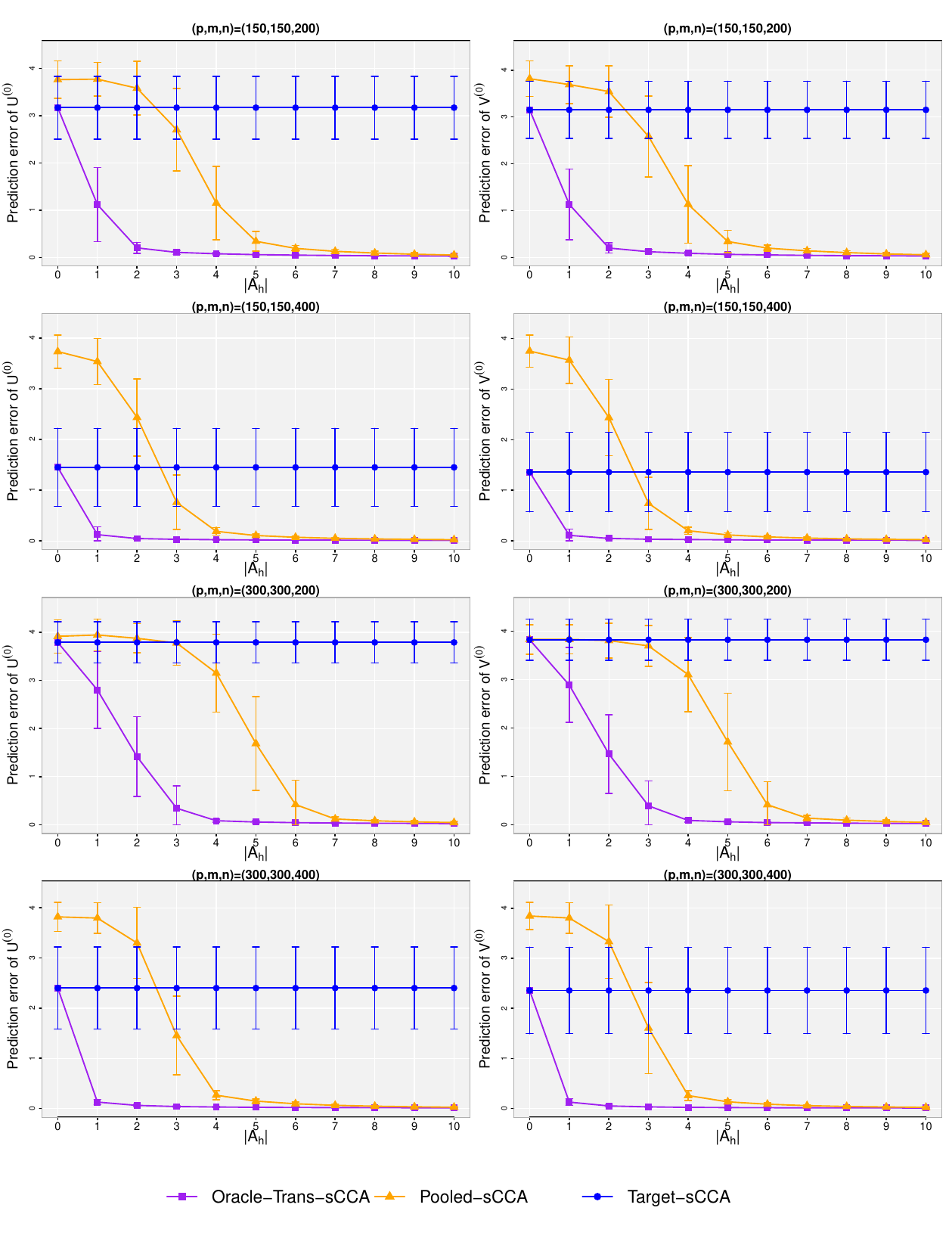}
    \captionsetup{font=footnotesize}
    \caption{Simulation results for the exactly low-rank setting. Each row corresponds to a different choice of $(p,m,n)$. The left and right columns report the prediction errors for $\boldsymbol{U}^{(0)}$ and $\boldsymbol{V}^{(0)}$, respectively, averaged over $100$ replications. Error bars denote the standard deviations.}
    \label{fig1}
\end{figure}

\begin{figure}[!ht]
    \centering
    \includegraphics[width=\textwidth]{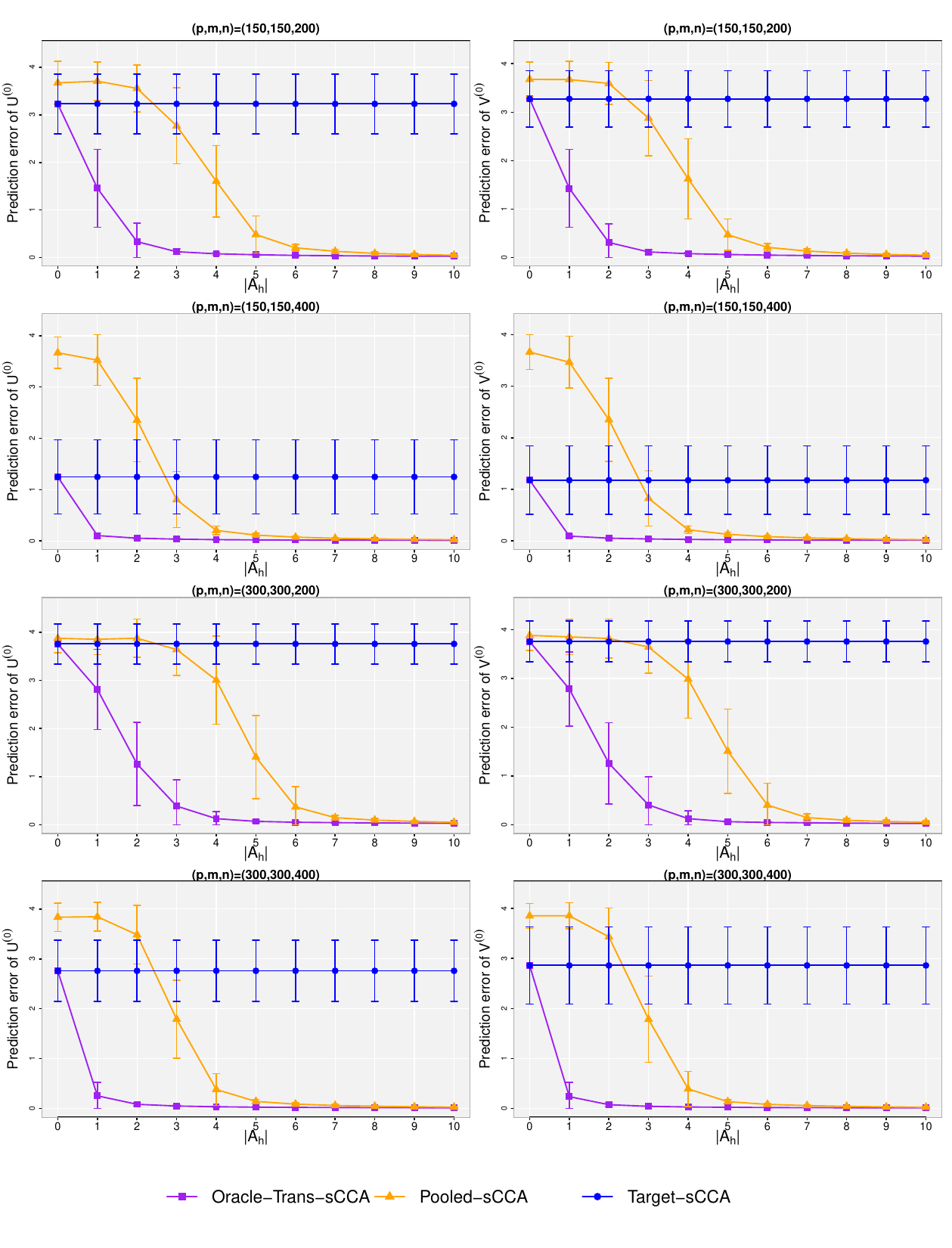}
    \captionsetup{font=footnotesize}
    \caption{Simulation results for the approximately low-rank setting. Each row corresponds to a different choice of $(p,m,n)$. The left and right columns report the prediction errors for $\boldsymbol{U}^{(0)}$ and $\boldsymbol{V}^{(0)}$, respectively, averaged over $100$ replications. Error bars denote the standard deviations.}
    \label{fig2}
\end{figure}

We evaluate coefficient estimation using the prediction loss (2.5) and its symmetric counterpart for $\boldsymbol{V}^{(0)}$, computed from each method's final output without additional renormalization. The results are exhibited in Figures~\ref{fig1}--\ref{fig2}.

From Figures \ref{fig1} and \ref{fig2}, we draw several conclusions. First, across all considered dimensional and sample-size settings, \texttt{Oracle-Trans-sCCA} substantially improves upon \texttt{Target-sCCA} once informative source datasets are available. Moreover, its prediction errors for both $\boldsymbol{U}^{(0)}$ and $\boldsymbol{V}^{(0)}$ decrease rapidly as $|\mathcal{A}_h|$ increases, which is consistent with the theoretical gain from the enlarged effective sample size provided by informative sources. Second, \texttt{Pooled-sCCA} can perform substantially worse than \texttt{Oracle-Trans-sCCA}, and even worse than \texttt{Target-sCCA} when only a small proportion of the source datasets are informative. This demonstrates that indiscriminately pooling heterogeneous source datasets may introduce considerable transfer bias and lead to negative transfer. As the number of informative sources increases, the performance of \texttt{Pooled-sCCA} gradually improves and eventually approaches that of \texttt{Oracle-Trans-sCCA}, reflecting the diminishing impact of non-informative sources. Third, similar results are observed under both the exact and approximate settings and for the estimation of both canonical coefficient matrices. This suggests that our estimator is robust to model misspecification arising when the number of canonical correlations exceeds $2$. Overall, these results demonstrate the potential efficiency gain from transferring information across related datasets, while also highlighting the importance of identifying informative sources rather than naively pooling all available source data.

We further conduct a second set of simulations in which the informative source set is unknown and is first identified using the source detection procedure developed in Section \ref{sec4.2}. The detected source set is then used as the input to Algorithm \ref{algo2}. The corresponding simulation results are reported in Section C of the Supplementary Material.

\subsection{A Real Data Analysis}\label{sec5.2}
We evaluate the proposed method using multi-omics data from The Cancer Genome Atlas (TCGA). The data are available from the R Bioconductor package \texttt{curatedTCGAData} (\url{https://bioconductor.org/packages/curatedTCGAData}) \citep{ramos2020multiomic}. This package provides curated TCGA molecular profiles organized as \texttt{MultiAssayExperiment} objects, enabling the integration of multiple assays and matched patient identifiers \citep{ramos2017software}.

We consider five cancer cohorts, including esophageal carcinoma (ESCA), lung adenocarcinoma (LUAD), lung squamous cell carcinoma (LUSC), head and neck squamous cell carcinoma (HNSC), and bladder urothelial carcinoma (BLCA). For each cohort, we extract paired RNA-sequencing gene expression profiles and reverse-phase protein array (RPPA) measurements from the same patients. The RNA-seq data contain gene-level expression measurements, whereas the RPPA data quantify protein abundance levels. After matching patients across the two assays and removing unmatched samples, we obtain paired multi-omics datasets for subsequent analysis.

We view ESCA as the target dataset, and the other four cancer cohorts as the source datasets. To reduce the dimensionality of gene expression data, we perform screening for ESCA by retaining the top $2\%$ genes with the largest empirical variances in the target cohort and apply the selected gene set to all auxiliary datasets. Missing values in both omics platforms are imputed separately within each cohort using $k$-nearest-neighbor imputation with $k=10$. After preprocessing, the target dataset consists of $n^{(0)}=125$ paired samples, with $p=357$ RNA-seq gene expression features and $m=136$ RPPA protein abundance features. The auxiliary datasets contain the same $p$ RNA-seq features and $m$ RPPA features, with sample sizes ranging from $212$ to $360$. 

To evaluate the estimated canonical directions, we assess their out-of-sample latent correlations. Specifically, for each random split, we randomly select $80\%$ of the samples as the training set and use the remaining $20\%$ samples for evaluation. The canonical coefficient matrices are estimated using only the training samples, and the held-out samples are projected onto the estimated canonical spaces. The canonical correlations between the projected RNA-seq and RPPA measurements are then calculated on the test samples.

We compare our method \texttt{Trans-sCCA} with \texttt{Target-sCCA} \citep{gao2017sparse} and \texttt{Pooled-sCCA} by setting $r=2$. For \texttt{Trans-sCCA}, we first apply the source detection procedure in Section \ref{sec4.2} to obtain the estimated informative set $\widehat{\mathcal{A}}_{\hat{q}}$, and then apply Algorithm \ref{algo2} with $\widehat{\mathcal{A}}_{\hat{q}}$ as input. Here, although Algorithm \ref{algo2} is formulated with a three-way sample splitting scheme to facilitate the theoretical analysis, in the practical implementation we do not further split the training samples. Instead, the entire target training sample, together with all available source samples, is used at each stage of the estimation procedure. Each experiment is repeated over $50$ random train-test splits. We report the first two canonical correlations (CC1 and CC2), and their sum (Total CC) in the following table.

\begin{table*}[htbp]
	\caption{Comparison of canonical correlations for ESCA. The empirical average (and standard deviation) of canonical correlations, over $50$ replications, are reported.}
	\resizebox{\linewidth}{!}{
		\centering
		\begin{tabular}{ccccccccccccc}
			\hline
			Methods&CC1&CC2&Total CC\cr
            \texttt{Trans-sCCA}& $0.8523(0.0586)$&$0.2559(0.1902)$&$1.1082(0.2142)$\cr
            \texttt{Pooled-sCCA}& $0.7870(0.0486)$&$0.1806(0.1373)$&$0.9675(0.1293)$\cr
            \texttt{Target-sCCA}& $0.7110(0.1618)$&$0.2332(0.1291)$&$0.9442(0.2249)$\cr
			\hline
		\end{tabular}
	}
	\label{tab1}
\end{table*}

Table~\ref{tab1} shows that \texttt{Trans-sCCA} achieves the largest out-of-sample canonical correlations when ESCA is treated as the target dataset. In particular, it yields a Total CC of $1.1082$, compared with $0.9675$ for \texttt{Pooled-sCCA} and $0.9442$ for \texttt{Target-sCCA}. The improvement is mainly driven by a substantially larger first canonical correlation, while a similar advantage is also observed for the second component. These results suggest that selectively incorporating informative source datasets can improve the estimation of the canonical structure for the target population.

\begin{figure}[!ht]
    \centering
    \includegraphics[width=\textwidth]{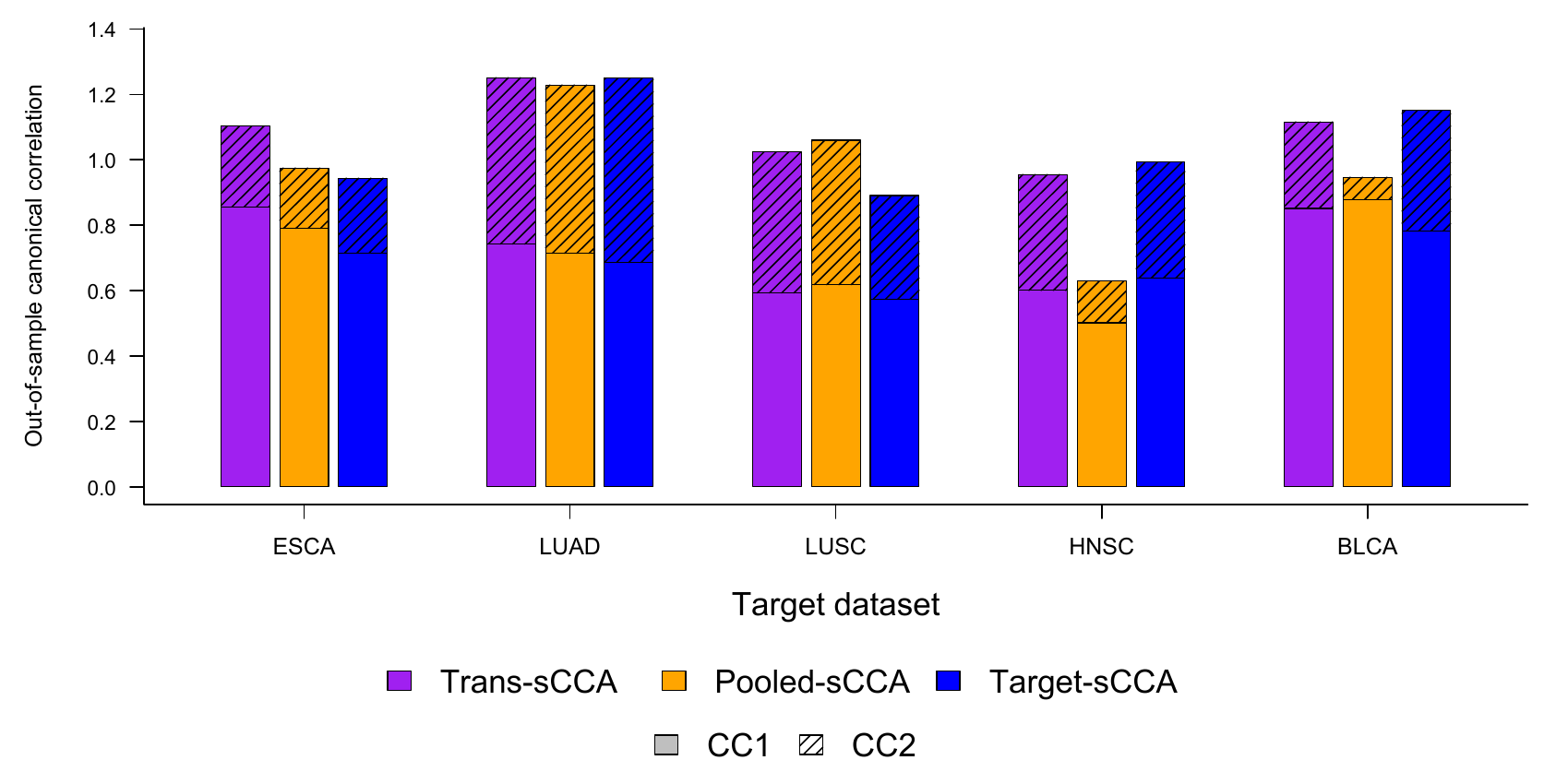}
    \captionsetup{font=footnotesize}
    \caption{Out-of-sample canonical correlations across five target cancer cohorts. For each target cohort, the three bars correspond to \texttt{Trans-sCCA}, \texttt{Pooled-sCCA}, and \texttt{Target-sCCA}. The solid and hatched portions represent the first and second canonical correlations (CC1 and CC2), respectively, so that the total height of each bar represents Total CC.}
    \label{fig5}
\end{figure}

To further examine the performance across different target populations, we repeat the analysis by treating each of the five cancer cohorts in turn as the target dataset, with the remaining cohorts serving as auxiliary sources. Figure~\ref{fig5} summarizes the resulting CC1 and CC2, with the total height of each bar representing Total CC. Relative to \texttt{Pooled-sCCA}, \texttt{Trans-sCCA} is less affected by heterogeneous source datasets, while still benefiting from informative auxiliary samples when sufficient source-target similarity is present. Across the five target cohorts, \texttt{Trans-sCCA} shows consistently competitive performance, with a particularly pronounced improvement for ESCA. Overall, these results indicate that the proposed data-driven procedure can borrow useful information from related sources while limiting the adverse effect of non-informative ones.

\section{Conclusion}\label{sec6}
In this paper, we developed a transfer learning framework for high-dimensional sparse CCA. By measuring source-target similarity through the spectral discrepancy between joint covariance matrices, we established minimax rates under both joint and separate losses, thereby characterizing the gain from informative source data and the price of heterogeneity. We proposed statistically optimal procedures that combine transfer, debiasing/aggregation, and truncated normalization, and showed that they attain the minimax rates in the sense of prediction loss under suitable conditions. A generalized sin-theta theorem was also developed to support the perturbation analysis without requiring comparability between the leading and smallest signal singular values. These results provide a theoretical foundation for transfer learning in sparse CCA and demonstrate when and how auxiliary datasets can improve estimation of the target canonical structure. Extensions to rank determination, non-Gaussian or nonlinear settings, and differential privacy are promising directions for future research.

\begin{acks}[Acknowledgments]
The authors were supported by NSFC under Grant No. 12571311.
\end{acks}



\bibliographystyle{imsart-nameyear}
\bibliography{bibliography.bib}

@article{gao2017sparse,
  title={Sparse CCA: adaptive estimation and computational barriers},
  author={Gao, Chao and Ma, Zongming and Zhou, Harrison H},
  journal={The Annals of Statistics},
  volume={45},
  number={5},
  pages={2074},
  year={2017},
  publisher={Institute of Mathematical Statistics}
}

@article{gao2015minimax,
  title={Minimax estimation in sparse canonical correlation analysis},
  author={Gao, Chao and Ma, Zongming and Ren, Zhao and Zhou, Harrison H},
  journal={The Annals of Statistics},
  pages={2168--2197},
  year={2015},
  publisher={JSTOR}
}

@article{hotelling1936relations,
  title={Relations between two sets of variates},
  author={Hotelling, Harold},
  journal={Biometrika},
  volume={28},
  number={3-4},
  pages={321--377},
  year={1936},
  publisher={JSTOR}
}

@article{bao2019canonical,
  title={Canonical correlation coefficients of high-dimensional Gaussian vectors: Finite rank case},
  author={Bao, Zhigang and Hu, Jiang and Pan, Guangming and Zhou, Wang},
  journal={The Annals of Statistics},
  volume={47},
  number={1},
  pages={612--640},
  year={2019},
  publisher={JSTOR}
}

@incollection{bai2022limiting,
  title={Limiting canonical distribution of two large-dimensional random vectors},
  author={Bai, Zhidong and Hou, Zhiqiang and Hu, Jiang and Jiang, Dandan and Zhang, Xiaozhuo},
  booktitle={Methodology and Applications of Statistics: A Volume in Honor of CR Rao on the Occasion of his 100th Birthday},
  pages={213--238},
  year={2022},
  publisher={Springer}
}

@article{hardoon2011sparse,
  title={Sparse canonical correlation analysis},
  author={Hardoon, David R and Shawe-Taylor, John},
  journal={Machine Learning},
  volume={83},
  number={3},
  pages={331--353},
  year={2011},
  publisher={Springer}
}

@article{le2009sparse,
  title={Sparse canonical methods for biological data integration: application to a cross-platform study},
  author={L{\^e} Cao, Kim-Anh and Martin, Pascal GP and Robert-Grani{\'e}, Christ{\`e}le and Besse, Philippe},
  journal={BMC bioinformatics},
  volume={10},
  number={1},
  pages={34},
  year={2009},
  publisher={Springer}
}

@article{parkhomenko2009sparse,
  title={Sparse Canonical Correlation Analysis with Application to Genomic Data Integration},
  author={Parkhomenko, Elena and Tritchler, David and Beyene, Joseph},
  journal={Statistical Applications in Genetics and Molecular Biology},
  volume={8},
  number={1},
  pages={36},
  year={2009},
  publisher={Walter de Gruyter GmbH}
}

@article{waaijenborg2009sparse,
  title={Sparse canonical correlation analysis for identifying, connecting and completing gene-expression networks},
  author={Waaijenborg, Sandra and Zwinderman, Aeilko H},
  journal={BMC bioinformatics},
  volume={10},
  number={1},
  pages={315},
  year={2009},
  publisher={Springer}
}

@article{witten2009penalized,
  title={A penalized matrix decomposition, with applications to sparse principal components and canonical correlation analysis},
  author={Witten, Daniela M and Tibshirani, Robert and Hastie, Trevor},
  journal={Biostatistics},
  volume={10},
  number={3},
  pages={515--534},
  year={2009},
  publisher={Oxford University Press}
}

@article{chen2013sparse,
  title={Sparse CCA via precision adjusted iterative thresholding},
  author={Chen, Mengjie and Gao, Chao and Ren, Zhao and Zhou, Harrison H},
  journal={arXiv preprint arXiv:1311.6186},
  year={2013}
}

@article{helmer2024stability,
  title={On the stability of canonical correlation analysis and partial least squares with application to brain-behavior associations},
  author={Helmer, Markus and Warrington, Shaun and Mohammadi-Nejad, Ali-Reza and Ji, Jie Lisa and Howell, Amber and Rosand, Benjamin and Anticevic, Alan and Sotiropoulos, Stamatios N and Murray, John D},
  journal={Communications biology},
  volume={7},
  number={1},
  pages={217},
  year={2024},
  publisher={Nature Publishing Group UK London}
}

@article{jiang2023canonical,
  title={Canonical correlation analysis for multi-omics: application to cross-cohort analysis},
  author={Jiang, Min-Zhi and Aguet, Fran{\c{c}}ois and Ardlie, Kristin and Chen, Jiawen and Cornell, Elaine and Cruz, Dan and Durda, Peter and Gabriel, Stacey B and Gerszten, Robert E and Guo, Xiuqing and others},
  journal={PLoS genetics},
  volume={19},
  number={5},
  pages={e1010517},
  year={2023},
  publisher={Public Library of Science San Francisco, CA USA}
}

@article{butler2018integrating,
  title={Integrating single-cell transcriptomic data across different conditions, technologies, and species},
  author={Butler, Andrew and Hoffman, Paul and Smibert, Peter and Papalexi, Efthymia and Satija, Rahul},
  journal={Nature biotechnology},
  volume={36},
  number={5},
  pages={411--420},
  year={2018},
  publisher={Nature Publishing Group}
}

@article{li2024knowledge,
  title={Knowledge transfer across multiple principal component analysis studies},
  author={Li, Zeyu and Qin, Kangxiang and He, Yong and Zhou, Wang and Zhang, Xinsheng},
  journal={arXiv preprint arXiv:2403.07431},
  year={2024}
}

@article{cai2026transfer,
  title={Transfer learning for covariance matrix estimation: optimality and adaptivity},
  author={Cai, T Tony and Kim, Dongwoo and Li, Yicheng and Xiang, Dongdong},
  journal={Journal of the American Statistical Association},
  number={just-accepted},
  pages={1--56},
  year={2026},
  publisher={Taylor \& Francis}
}

@article{yuan2006model,
  title={Model selection and estimation in regression with grouped variables},
  author={Yuan, Ming and Lin, Yi},
  journal={Journal of the Royal Statistical Society Series B: Statistical Methodology},
  volume={68},
  number={1},
  pages={49--67},
  year={2006},
  publisher={Oxford University Press}
}

@article{zhao2026trans,
  title={Trans-glasso: A transfer learning approach to precision matrix estimation},
  author={Zhao, Boxin and Ma, Cong and Kolar, Mladen},
  journal={Journal of the American Statistical Association},
  pages={1--13},
  year={2026},
  publisher={Taylor \& Francis}
}

@article{li2023transfer,
  title={Transfer learning in large-scale gaussian graphical models with false discovery rate control},
  author={Li, Sai and Cai, T Tony and Li, Hongzhe},
  journal={Journal of the American Statistical Association},
  volume={118},
  number={543},
  pages={2171--2183},
  year={2023},
  publisher={Taylor \& Francis}
}

@article{pan2009survey,
  title={A survey on transfer learning},
  author={Pan, Sinno Jialin and Yang, Qiang},
  journal={IEEE Transactions on knowledge and data engineering},
  volume={22},
  number={10},
  pages={1345--1359},
  year={2009},
  publisher={IEEE}
}

@article{jin2024transfer,
  title={Transfer learning with large-scale quantile regression},
  author={Jin, Jun and Yan, Jun and Aseltine, Robert H and Chen, Kun},
  journal={Technometrics},
  volume={66},
  number={3},
  pages={381--393},
  year={2024},
  publisher={Taylor \& Francis}
}

@article{li2024estimation,
  title={Estimation and inference for high-dimensional generalized linear models with knowledge transfer},
  author={Li, Sai and Zhang, Linjun and Cai, T Tony and Li, Hongzhe},
  journal={Journal of the American Statistical Association},
  volume={119},
  number={546},
  pages={1274--1285},
  year={2024},
  publisher={Taylor \& Francis}
}

@article{fu2025transfer,
  title={Transfer learning for high-dimensional Factor-augmented sparse linear model},
  author={Fu, Bo and Jiang, Dandan},
  journal={arXiv preprint arXiv:2511.12435},
  year={2025}
}

@article{cai2021transfer,
  title={Transfer learning for nonparametric classification},
  author={Cai, T Tony and Wei, Hongji},
  journal={The Annals of Statistics},
  volume={49},
  number={1},
  pages={100--128},
  year={2021},
  publisher={JSTOR}
}

@article{li2022transfer,
  title={Transfer learning for high-dimensional linear regression: Prediction, estimation and minimax optimality},
  author={Li, Sai and Cai, T Tony and Li, Hongzhe},
  journal={Journal of the Royal Statistical Society: Series B (Statistical Methodology)},
  volume={84},
  number={1},
  pages={149--173},
  year={2022},
  publisher={Oxford University Press}
}

@article{tian2023transfer,
  title={Transfer learning under high-dimensional generalized linear models},
  author={Tian, Ye and Feng, Yang},
  journal={Journal of the American Statistical Association},
  volume={118},
  number={544},
  pages={2684--2697},
  year={2023},
  publisher={Taylor \& Francis}
}

@article{bastani2021predicting,
  title={Predicting with proxies: Transfer learning in high dimension},
  author={Bastani, Hamsa},
  journal={Management Science},
  volume={67},
  number={5},
  pages={2964--2984},
  year={2021},
  publisher={INFORMS}
}

@article{chai2025deep,
  title={Deep transfer $ Q $-learning for offline non-stationary reinforcement Learning},
  author={Chai, Jinhang and Chen, Elynn and Fan, Jianqing},
  journal={arXiv preprint arXiv:2501.04870},
  year={2025}
}

@article{hanneke2022no,
  title={A no-free-lunch theorem for multitask learning},
  author={Hanneke, Steve and Kpotufe, Samory},
  journal={The Annals of Statistics},
  volume={50},
  number={6},
  pages={3119--3143},
  year={2022},
  publisher={Institute of Mathematical Statistics}
}

@inproceedings{
luo2026fedcca,
title={Fed{CCA}: Federated Canonical Correlation Analysis},
author={Zhengquan Luo and Kai Fong Ernest Chong and Pengfei Wei and Changyou Chen and Peilin Zhao and Renmin Han and Chunlai Zhou and Yunlong Wang and zhiqiang xu},
booktitle={The 29th International Conference on Artificial Intelligence and Statistics},
year={2026},
url={https://openreview.net/forum?id=C1IA9PNO0Q}
}

@ARTICLE{7120160,
  author={Bertrand, Alexander and Moonen, Marc},
  journal={IEEE Transactions on Signal Processing}, 
  title={Distributed Canonical Correlation Analysis in Wireless Sensor Networks With Application to Distributed Blind Source Separation}, 
  year={2015},
  volume={63},
  number={18},
  pages={4800-4813},
  doi={10.1109/TSP.2015.2443729}}

@article{chen2024distributed,
  title={Distributed estimation and gap-free analysis of canonical correlations},
  author={Chen, Canyi and Zhu, Liping},
  journal={arXiv preprint arXiv:2412.17792},
  year={2024}
}

@article{mai2019iterative,
  title={An iterative penalized least squares approach to sparse canonical correlation analysis},
  author={Mai, Qing and Zhang, Xin},
  journal={Biometrics},
  volume={75},
  number={3},
  pages={734--744},
  year={2019},
  publisher={Oxford University Press}
}

@article{yoon2020sparse,
  title={Sparse semiparametric canonical correlation analysis for data of mixed types},
  author={Yoon, Grace and Carroll, Raymond J and Gaynanova, Irina},
  journal={Biometrika},
  volume={107},
  number={3},
  pages={609--625},
  year={2020},
  publisher={Oxford University Press}
}

@article{di2026sparse,
  title={Sparse Gaussianized Canonical Correlation Analysis with Applications to Portfolio Analysis},
  author={Di, He and Zou, Hui},
  journal={Journal of the American Statistical Association},
  pages={1--14},
  year={2026},
  publisher={Taylor \& Francis}
}

@article{argyriou2012sparse,
  title={Sparse prediction with the $ k $-support norm},
  author={Argyriou, Andreas and Foygel, Rina and Srebro, Nathan},
  journal={Advances in Neural Information Processing Systems},
  volume={25},
  year={2012}
}

@article{ramos2020multiomic,
  title={Multiomic integration of public oncology databases in bioconductor},
  author={Ramos, Marcel and Geistlinger, Ludwig and Oh, Sehyun and Schiffer, Lucas and Azhar, Rimsha and Kodali, Hanish and de Bruijn, Ino and Gao, Jianjiong and Carey, Vincent J and Morgan, Martin and others},
  journal={JCO clinical cancer informatics},
  volume={1},
  pages={958--971},
  year={2020},
  publisher={American Society of Clinical Oncology}
}

@article{ramos2017software,
  title={Software for the integration of multiomics experiments in bioconductor},
  author={Ramos, Marcel and Schiffer, Lucas and Re, Angela and Azhar, Rimsha and Basunia, Azfar and Rodriguez, Carmen and Chan, Tiffany and Chapman, Phil and Davis, Sean R and Gomez-Cabrero, David and others},
  journal={Cancer research},
  volume={77},
  number={21},
  pages={e39--e42},
  year={2017},
  publisher={American Association for Cancer Research}
}

\end{document}